\documentclass[%
 aip,
 amsmath,amssymb,
reprint,%
]{revtex4-1}
\usepackage[utf8]{inputenc}
\usepackage[T1]{fontenc}
\usepackage{mathptmx}
\usepackage{etoolbox}
\usepackage{lipsum}
\usepackage{graphicx}
\usepackage{amssymb}
\usepackage{url}
\usepackage{soul}
\usepackage{xcolor}
\usepackage[labelformat=simple]{subcaption}

\usepackage{float}
\usepackage{calligra}
\usepackage{calrsfs}
\usepackage{xcolor}
\usepackage{enumitem}
\usepackage{graphicx}
\usepackage{dcolumn}
\usepackage{bm}
\usepackage{tabularx,ragged2e,booktabs,caption}
\usepackage{xcolor}
\usepackage[left=2cm,right=2cm,top=2cm,bottom=2cm]{geometry}

\definecolor{LWcolor}{RGB}{0,0,255} 

\newcommand{\mycomment}[1]{}
\usepackage{icomma}

\setcitestyle{numbers,square}

\definecolor{HGcolor}{RGB}{255,20,147} 

\definecolor{GScolor}{RGB}{0,0,0} 

\makeatletter
\def\@email#1#2{%
 \endgroup
 \patchcmd{\titleblock@produce}
  {\frontmatter@RRAPformat}
  {\frontmatter@RRAPformat{\produce@RRAP{*#1\href{mailto:#2}{#2}}}\frontmatter@RRAPformat}
  {}{}
}%
\makeatother
\begin{document}
\title{{Large Eddy Simulations of Turbulent Pipe Flows at Moderate Reynolds Numbers}}
\author{Himani Garg}
\author{Lei Wang}%
\author{Martin Andersson}%
\author{Christer Fureby}
\email{himani.garg@energy.lth.se; himani.garg65@gmail.com}
 \affiliation{$^1$Lund University$,$ Department of Energy Sciences $,$ P$.$O$.$ Box 118$,$ SE-22100 Lund$,$ Sweden}
 
\begin{abstract}
\section*{Abstract}
Wall-bounded turbulence is relevant for many engineering and natural science applications, yet there are still aspects of its underlying physics that are not fully understood, particularly at high Reynolds numbers. In this study, we investigate fully-developed turbulent pipe flows at moderate-to-high friction velocity Reynolds numbers ($361 \leq Re_\tau \leq 2,000$), {corresponding to bulk velocity-based Reynolds numbers of $11,700 \leq Re_b \leq 82,500$}, using wall-modeled Large Eddy Simulations (LES) in OpenFOAM. A grid convergence study is performed for $Re_\tau=361$, followed by an investigation of the accuracy of various subgrid-scale stress models for the same Reynolds number. Results show that the Wall-Adapting Local Eddy (WALE) model performs well compared to experiments and Direct Numerical Simulations (DNS), while One-Equation Eddy-Viscosity Model (OEEVM) and Smagorinsky (SMG) are too dissipative. LES utilizing WALE are then performed for four different Reynolds numbers with gradually refined grids, revealing excellent agreement with DNS data in the outer region. However, a significant deviation from DNS data is observed in the sub-viscous layer region, indicating the need for further mesh refinement in the wall-normal direction to accurately capture the smallest-scale motions' behavior. Additional mesh sensitivity analysis uncovered that, as the $Re_\tau$ value rises, it becomes crucial for a grid to adhere to the condition of $\Delta x^+ \leq 20-25$ and $\Delta z^+ \leq 10$ in order to precisely capture substantial large and small scale fluctuations. Overall, the WALE model enables accurate numerical simulations of high-Reynolds-number, wall-bounded flows at a fraction of the computational cost required for temporal and spatial resolution of the inner layer.
\end{abstract}


\maketitle

\section{Introduction}\label{sec1}
Two common features of engineering and environmental fluid flows are wall-boundedness \citep{stull2000meteorology,thorpe2005turbulent} and high Reynolds number ($Re$) turbulence \citep{marusic2010wall,manneville2016transition,smits2011high}. The study of turbulent flow traces back over a century, and its widespread application has kept this field under constant development \citep{monty2005developments}. The spatially evolving boundary layer, the channel, and the pipe are three canonical building block flows for wall bounded turbulent flows. Consequently, pipe flow is fully described by $Re$ and the length of the pipe. The effect of the pipe length is insignificant only if considered sufficiently large \citep{el2013direct}. During the transport of fluids through pipes, roughly half of the energy is dissipated by turbulence near the walls \citep{jimenez2012cascades}. The need to investigate fundamental wall-bounded turbulence at high $Re$ is widely agreed in literature. Clauser \citep{clauser1956turbulent} and Coles \& Hirst \citep{coles1969computation} provided a comprehensive review of classical wall-bounded turbulent scaling and structure for the case of a smooth wall. However, one of the most recent revealing research result comes from efforts seeking to achieve an ever-higher $Re$ number and, in doing so, shedding light on some of the longstanding questions about the very nature of wall-bounded turbulence, including how well our present analytical and computational models match its complex behavior \citep{marusic2010wall}.

To date, accurate high-$Re$-number data are still primarily achievable only via well-designed experiments~\citep{2017CICLope}. With the progress of computational technology, simulations have emerged to play a vital role in research on turbulent flows and wall-bounded turbulence. One of the most cited articles on turbulent pipe flow is by Eggels \textit{et al.} \citep{eggels1994fully}, showing comparisons between Direct Numerical Simulation (DNS) and experimental results of Hot Wire Anemometry, Particle Image Velocimetry, and Laser Doppler Anemometry at nominally similar friction-velocity-based turbulent $Re_\tau = u_\tau h/\nu \simeq 180$, where $u_\tau$ the friction velocity, $h$ the characteristic length, and $\nu$ the kinematic viscosity. Numerous studies (e.g., \citep{antonia_teitel_kim_browne_1992, Toonder1997, WAGNER2001581, peng2018direct}) have extensively documented the effects at low $Re_\tau$. These studies illustrate the augmentation in the maxima of streamwise turbulence intensity and Reynolds shear stress, along with inner-scaled turbulence statistics correlated with $Re_\tau$. These effects are predominantly observed in the logarithmic law of the mean velocity profiles, attributed to higher pressure gradients. Nevertheless, decades of experimental research have shown that obtaining detailed high-$Re_\tau$ data, notably near the wall, remains challenging. That is because as $Re_\tau$ increases, the smallest length scales decrease, and as a result lead to significant uncertainties in determining the probe locations and turbulence intensities. Recently, DNS of turbulent pipe flows with $Re_\tau$ ranging from $500$ to $6,000$ have been presented \citep{wu_moin_2008,Chin2010,Klewicki2012,Lee2013,el2013direct,CHIN201433,Junsun2015,pirozzoli2021one,2021JieYao}. These $Re_\tau$ are considered modest compared to turbulent pipe flow experiments carried out with a turbulent $Re$ range of $Re_\tau \simeq O(10^3 - 10^5)$ \citep{guala_hommema_adrian_2006,monty_stewart_williams_chong_2007,zagarola1998mean,2017CICLope,furuichi2019experimental}.
  
As $Re_\tau$ increases, the range of turbulent scales extends, demanding higher resolutions for DNS. However, the number of required grid points for DNS scales roughly as $O(Re_\tau^{9/4} )$ \citep{coleman2010primer}, presently falling short of computing capability at engineering $Re_\tau$ values. 
Capturing the turbulent flow and accounting for additional computing overhead requires computer resources that scale nominally as $Re_\tau^4$ \citep{yakhot2005anomalous}. Considering that many real-world flows have $Re_\tau > 10^6$, the computing costs associated with DNS quickly becomes prohibitive. On the other hand Large Eddy Simulations (LES) takes a different approach, by filtering of the smallest scales and replacing their effects on the large resolved scales by subgrid models~\citep{sagaut2006large}. LES of unbounded homogeneous and shear turbulence is well established \citep{misra1997vortex,pitsch2005large,matheou2010large,ferrante2011inclined}, whereas LES of wall-bounded flows has remained challenging, firstly, because the presence of a wall causes the length scales to progressively decrease toward the wall, and secondly, owing to the near-wall anisotropy. 

When a wall is present the LES grid is typically constrained by the near-wall region, where turbulent and viscous momentum transport occur on small scales. Piomelli \& Balars \citep{piomelli2002wall} showed that for near-wall-resolved LES, around 99\% of grid points are used to resolve the wall layers in fully-developed channel flow. As a result, for wall-bounded flows the computational cost of near-wall-resolved LES to resolve the viscous sublayer is estimated to scale as $Re_\tau^{1.8}$, which demonstrates the potential savings over DNS \citep{piomelli2002wall}. Nevertheless, the development of wall-modeled LES \citep{piomelli2002wall,templeton2005eddy,piomelli2008wall,templeton2008predictive,chung2009large} has widened the range of applications of LES for wall-bounded flows. In near-wall-modeled LES, the effects of the near-wall anisotropic fine scales are not resolved but modeled by wall models, approximating the near-wall physics. Near-wall-modeled LES is computationally more efficient than near-wall-resolved LES since only the outer layer is explicitly computed, with the required grid size determined by the outer flow eddies \citep{piomelli2002wall,fureby2004large,georgiadis2010large,larsson2016large,fureby2017high}. Thus in wall-modeled LES, the grid requirements could be relaxed; it is estimated that the computational costs would scale as $Re_\tau^{0.2}$ \citep{smits2013wall}. 
Sagaut \citep{sagaut2006large} gives a thorough description of the (still on-going) efforts in the formulation of more accurate LES models. Yet few simulations have been performed at high $Re$.

To the best of our knowledge, the highest turbulent $Re_\tau$ used in LES of turbulent pipe flows were $Re_\tau \simeq 2,000$ and $180,000$, by Berrouk \textit{et al.} \citep{berrouk2007stochastic} and Ferrer \textit{et al.} \citep{ferrer2019high}. The authors present explicit and implicit LES of fully-developed turbulent pipe flows using a continuous Galerkin spectral element solver, motivated by the work of Chung \& Pullin \citep{chung2009large}. For explicit LES they utilized the explicit stretched-vortex model of Chung \& Pullin \citep{chung2009large}. From the literature review, it is evident that only a limited number of studies have been conducted with wall-modeled LES focusing on the moderate-to-high $Re_\tau$ wall-bounded turbulent flow. The objective of the present study is to carry out wall-modeled LES for moderate to high $Re_\tau$ and compare the results to DNS at matched $Re_\tau$ for internal wall-bounded flows. To perform wall-modeled LES, different meshes have been generated, with $y^+$ values from $4$ to $11$. We study the performance of different subgrid models, investing their accuracy at predicting the statistical behavior of moderate $Re$ fully-developed turbulent pipe flows. Five subgrid models are considered: the Wall-Adapting Local Eddy viscosity model (WALE) \citep{nicoud1999subgrid}, the Smagorinsky model (SMG) \citep{smagorinsky1963general}, the One-Equation Eddy Viscosity Model (OEEVM) \citep{schumann1975subgrid}, the Localized Dynamic Kinetic energy Model (LDKM) \citep{kim1995new,kim1997application}, and the Differential Stress Equation Model (DSEM) \citep{deardorff1973use}.  Only the WALE model is selected to investigate the dependence on grid resolution, as a trade-off between computational cost and numerical accuracy. For detailed investigation, four friction-velocity-based turbulent $Re$ are considered: $Re_\tau = 361$, $550$, $1,000$, and $2,000$. The simulations are evaluated based on predictions of zeroth-, first-, and second-order turbulence statistics and flow features such as friction coefficient, mean velocity profiles, velocity correlations, mean turbulent shear stress profiles, quadrant analysis, and anisotropy invariant maps. Numerical results for each model are compared with DNS data \citep{el2013direct,pirozzoli2021one}. 
\section{LES methodology}
\subsection{Mathematical formulation}
In LES, the governing equations are the low pass filtered conservation equations of incompressible flow mass and momentum as represented by,
\begin{equation}
\dfrac{\partial\overline{u}_i}{\partial x_i} = 0, \label{eq:2}
\end{equation}

\begin{equation}
\frac{\partial \overline{u}_i}{\partial t} + \frac{\partial}{\partial x_j}(\overline{u}_i\overline{u}_j) = -\frac{1}{\rho}\frac{\partial\overline{p}}{\partial x_i}-\frac{\partial\tau^{sgs}_{ij}}{\partial x_j}+\frac{\partial}{\partial x_j} \left(\nu \frac{\partial \overline{u}_i}{\partial x_j} \right), 
\label{eq:2bis}
\end{equation}
where $\overline{{u}}_{i}$ is the filtered velocity, $\overline{p}$, the filtered pressure, $\nu$, the kinematic viscosity, and $\tau^{\text{sgs}}_{ij} = \overline{u_i u_j}-\overline{u_i}\,\overline{u_j}$, the subgrid stress tensor. Mathematical analysis and physical arguments, e.g., \citep{speziale1985galilean,vreman1994realizability,fureby1997mathematical}, suggest that: (i) $\tau^{\text{sgs}}_{ij}$ is invariant under a change of frame; (ii) $\tau^{\text{sgs}}_{ij}$ is positive-definite symmetric, provided that the filter function, $G_\Delta$, is symmetric; and (iii) the inequalities $k=\frac{1}{2} \text{tr}(\tau^{\text{sgs}}_{ij}) \geq 0$, $k^2 \geq \left\| \tau^{\text{sgs}}_{ij} \right\|$, and det($\tau^{\text{sgs}}_{ij}$) $\geq 0$ must be satisfied for $\tau^{\text{sgs}}_{ij}$ to be positive-definite.
\subsection{Subgrid modeling}
Two methodologies are usually adopted for subgrid modeling: functional modeling, which consists in modeling the action of the subgrid scales on the resolved scales, and structural modeling, whose principle is to model the subgrid stresses without incorporating any knowledge about the interactions between the subgrid and resolved scales \citep{sagaut2006large}. However, for the purpose of this paper, we instead prefer to focus on isotropic and anisotropic subgrid models since complex flows with high $Re$ numbers are often characterized by anisotropy on a wide range of scales -- typically reaching the range of scales that require modeling, e.g., in the near-wall region. The most frequently used subgrid models are the eddy-viscosity models. This type of models computes the deviatoric part of the subgrid stress tensor by relating it, in a linear correspondence, to the rate-of-strain tensor, ${S}_{ij} = \frac{1}{2}(\partial {u}_i/ \partial x_j + \partial {u}_j/ \partial x_i)$~\citep{li2009matrix}. By doing so, the subgrid stress tensor may be thought of being decomposed into an isotropic part and an anisotropic part. The former is also known as the normal stresses and is directly related to the turbulent kinetic energy while the latter measures the deviations from isotropy. Hence, the subgrid stress tensor, based on the Boussinesq hypothesis, is expressed as
\begin{equation}
{\tau^{\text{sgs}}_{ij} - \frac{2}{3}k_{\text{sgs}}\delta_{ij} = -2 \nu_{\text{sgs}}\overline{S}_{ij},}
\label{eq:3}
\end{equation}
where $\nu_{\text{sgs}}$ is the subgrid eddy viscosity and $\delta_{ij}$ is the Kronecker delta. Models for the eddy viscosity, $\nu_{\text{sgs}}$, and the subgrid turbulent kinetic energy, $k_{\text{sgs}}=\frac{1}{2}\text{tr}(\tau^\text{sgs}_{ii})$, are needed to close Eq. \eqref{eq:3}. For this purpose, we assume the existence of characteristic length and velocity scales and we infer separation between the resolved and subgrid scales. 

In this work, we limit ourselves to five well-known LES subgrid models: Firstly, the One-Equation Eddy Viscosity Model (OEEVM) \citep{schumann1975subgrid}, in which a modeled transport equation for $k_{\text{sgs}}$ is solved, and 
$\nu_{\text{sgs}} = c_k\Delta k_{\text{sgs}}^{1/2}$ is modeled, where $c_k=0.094$ and $c_\epsilon = 1.048$.

The second model is the Smagorinsky (SMG) model \citep{smagorinsky1963general} which assumes a local equilibrium between production and dissipation of $k_{sgs}$. As a result, the subgrid viscosity and kinetic energy becomes, $\nu_{\text{sgs}}=c_s^2 \Delta^2 (2\overline{S}_{ij}\overline{S}_{ij})^{1/2}$ and  $k_{\text {sgs}}= c_I\Delta^2(2\overline{S}_{ij}\overline{S}_{ij})$, where $c_k=0.094, c_\epsilon=1.048, c_I = c_k/c_\epsilon=0.089 ,$ and $c_s=c_k\sqrt{c_k/c_\epsilon}$ are the model coefficients~\citep{smagorinsky1963general,lilly1966application}. 

{The third} model is the Localized Dynamic Kinetic energy Model (LDKM) based on the work of Kim and Menon \citep{kim1995new,kim1997application}. In the LDKM, scale similarity between the subgrid-scale stress tensor and the test-scale Leonard stress tensor is assumed to evaluate the model coefficient{s, for $\nu_{sgs}$ and the subgrid dissipation, $\epsilon$}. The transport equation used in this model is similar to that in OEEVM model. The expression for $\nu_{sgs}$ is similar to the one of the OEEVM model: {$\nu_{{sgs}} = c_k\Delta k_{\text{sgs}}^{1/2}$}, except that the model coefficient, $c_k$, is now evaluated dynamically as
$c_k= \dfrac{1}{2}\dfrac{L_{ij}M_{ij}}{M_{ij}M_{ij}}$, where 
$M_{ij} = -\widetilde{\Delta}\textsc{k}^{1/2}\widetilde{\overline{S}}_{ij}$, $L_{ij} = \widetilde{\overline{u}_i\overline{u}_j} - \widetilde{\overline{u}_i}\widetilde{\overline{u}_j}$, $\textsc{k} = \frac{1}{2}L_{ii}$, and the tilde symbol, $\widetilde{(.)}$, denotes the explicit filtering operation, usually defined as twice the original filter, e.g., $\widetilde{\Delta} = 2\Delta$. {The subgrid dissipation is modeled as $\epsilon \approx c_\epsilon  \textsc{k}^{3/2}/\widetilde{\Delta}$ in which $c_\epsilon$ is evaluated dynamically as $c_\epsilon = \frac{\widetilde{\overline{g}_{ij}\overline{g}_{ij}}-\widetilde{\overline{g}_{ij}}\widetilde{\overline{g}_{ij}}}{\left[\left(\nu+\nu_{sgs}  \right)\widetilde{\Delta}\right]^{-1} \textsc{k}^{3/2}}$, with $\overline{g}_{ij} = \partial \overline{u}_i / \partial x_j$ being the filtered velocity gradient.}

The fourth model is the Wall-Adapting Local Eddy viscosity (WALE) model~\citep{nicoud1999subgrid}. This method allows to obtain a subgrid viscosity proportional to the wall-normal distance cubed ($\nu_\text{sgs} \propto y^3$), and additionally allows the subgrid viscosity to vanish near the solid walls. The WALE model can be defined via the following relation for subgrid viscosity,
$\nu_\text{sgs} = c_w\Delta^2 \frac{(S^d_{ij}S^d_{ij})^{3/2}}{(\overline{S}_{ij}\overline{S}_{ij})^{5/2} + (S^d_{ij}S^d_{ij})^{5/4}}$,
where $S^d_{ij} = \frac{1}{2}(\overline{g}^2_{ij}+\overline{g}^2_{ji}) - \frac{1}{3}\delta_{ij}\overline{g}^2_{ij}$ is the traceless symmetric part of the square of the velocity gradient tensor.

Lastly, we consider the Differential Stress Equation Model (DSEM) by Deardorff \citep{deardorff1973use}, which does not rely on the local isotropy assumption within the subgrid scales. 
The DSEM employed here uses a modeled transport equation for the subgrid stress tensor, $\tau_{ij}$, following \citep{deardorff1973use},
\begin{equation}
\begin{split}
\frac{\partial \tau_{ij}^\text{sgs}}{\partial t} =- \frac{\partial}{\partial x_k}(\tau_{ij}^\text{sgs}\overline{u}_{k}) -\tau_{ik}^\text{sgs}\frac{\partial \overline{u}_{j}}{\partial x_k}-\tau_{jk}^\text{sgs}\frac{\partial \overline{u}_{i}}{\partial x_k} \\
-\frac{\partial }{\partial x_k}\overline{u'_iu'_ju'_k} +p' \overline{\left(\frac{\partial u'_i}{\partial x_j} +\frac{\partial u'_j}{\partial x_i} \right)} -2\nu\overline{\frac{\partial u'_i}{\partial x_k} \frac{\partial u'_j}{\partial x_k}}.
\end{split}
\label{eq:11}
\end{equation}
The fourth, fifth, and sixth terms on the right-hand side of Eq.~\ref{eq:11}, which represent triple correlation, pressure-strain correlation, and the dissipation term, are modeled, and details can be found in \citep{fureby1997differential,sagaut2006large}.

The van Driest wall-damping function is used in the wall treatment for the SMG and OEEVM subgrid models. Whereas, a wall-model is instead used for the WALE, DSEM, and LDKM models. This wall model is based on the Spalding's law of the wall from which the friction velocity can be estimated based on which a wall viscosity can be derived at the first off-wall grid point, following the ideas of \citep{fureby2004large}.
\section{Simulation details}
The LES equations are solved numerically using the finite volume method with a second-order cell-centered discretization scheme using OpenFOAM. Time stepping is performed using the second-order accurate implicit Adams-Bashforth method \citep{butcher2016numerical} with a maximum Courant-Friedrichs-Lewy (CFL) number of 0.5 for numerical stability. The PISO algorithm for LES is used to decouple the pressure-momentum equations~\citep{weller1998tensorial}. The resulting pressure equation is solved using a Generalized Geometric-Algebraic Multigrid (GAMG) method with a Diagonal Incomplete Cholesky (DIC) smoother \citep{van1996matrix}. Velocity, turbulent kinetic energy, and dissipation rate are solved using a Preconditioned BiConjugate Gradient (PBiCG) solver \citep{press2007numerical} with a Diagonal Incomplete Lower-Upper (DILU) preconditioning method \citep{van1996matrix}. 

The turbulent bulk $Re_b$ number based on the pipe diameter, $D$, and bulk velocity, $U_b$, is given by
\begin{equation}
Re_b = \dfrac{U_bD}{\nu},
\end{equation}
yielding the four values $Re_b = 11,700$, $19,000$, $37,700$, and $82,500$ used in this study. {The motivation behind selecting these $Re_b$ values is the well-established DNS results previously reported \citep{el2013direct, pirozzoli2021one}. To provide a direct comparison of LES results with DNS, we use the same values of $Re_b$ as reported in these papers.} In addition to the bulk velocity, another characteristic velocity scale is commonly introduced in connection with wall-bounded flows. The friction velocity, $u_\tau$, is defined in terms of the wall shear stress, $\tau_w$, and the fluid density, $\rho$, as
\begin{equation}
u_\tau = \sqrt{\dfrac{\tau_w}{\rho}}.
\end{equation}
The turbulent friction $Re_\tau$ number, based on the pipe radius, $R$, and friction velocity, $u_\tau$, is given by
\begin{equation}
Re_\tau = \dfrac{u_\tau R}{\nu},
\end{equation}
yielding the four values $Re_\tau = 361$, $550$, $1,000$, and $2,000$ used in this study. 

Hexahedral butterfly grids, have been used for all simulations. These grids are composed of a Cartesian part at the center of the pipe surrounded by four transitional regions from the inner square to the outer circle of all cross-sections of the pipe. Such grids thus require the definition of a partition of the pipe into five blocks prior to the grid generation, but generally provide the best grid quality, i.e., high orthogonality and low grid density. The generation of these grids is automated by means of an m4 macro \citep{kernighan1977m4} in OpenFOAM. The length of the computational domain is $L$. The radial, axial, and azimuthal directions are represented by $r$, $x$, and $\theta$, and the corresponding velocity components, by $u$, $v$, and $w$. The wall-normal distance, i.e., along the radial direction, is denoted by $y = R-r$, where $R = D/2$ is the pipe radius. Here, $u'$, $v'$, and $w'$ are the velocity fluctuations in the streamwise, radial, and azimuthal directions, respectively. 

Since the fully-developed turbulent pipe flow considered here is homogeneous in the streamwise direction, periodic boundary conditions are imposed in this direction. No-slip boundary conditions are imposed at the pipe wall for all velocity components, whereas zero-Neumann boundary condition is used for the pressure. An additional external force term is introduced into the NSE to mimic the effect of the pressure gradient in the context of periodic boundary conditions in the streamwise direction. The magnitude of this force is determined by the bulk velocity. For the present study we chose the pipe length to be five times the pipe diameter, i.e., $L=10R$. 
Indeed, to ensure that statistics are not affected by the streamwise periodic boundary conditions, we have investigated one-dimensional temporal and spatial autocorrelation functions of velocity fluctuations. The results are presented in Appendix~\ref{Appendix I} and show that the chosen domain length is sufficiently large for velocity fluctuations to uncorrelate. {All flow statistics are averaged over approximately 100 to 120 flow-through times.}
\section{\label{sec:3}Results and discussions}
This discussion is mainly subdivided into three main sections. In Sec. \ref{Sec: MeshIndependence} simulations for statistically fully-developed turbulent pipe flow at $Re_\tau =361$ are presented. These simulations correspond to four different mesh resolutions. The results are grouped according to the computed quantites, e.g., mean flow profiles, Reynolds shear stresses, and friction factor. In Sec. \ref{Sec: SGS model dependence} fully-developed turbulent pipe flow simulations at fixed $Re_\tau =361$, for five LES models are performed.  In Sec. \ref{Sec: Re dependence} LES-WALE simulations for four moderate-to-high turbulent $Re_\tau$ are carried out. Again, in both sections results are grouped based on the computed quantities.  
\begin{table*}[!ht]
\caption{Mesh properties.}\label{table: Mesh information}%
\begin{tabular}{@{}llllll@{}}
\toprule
Mesh & Ncells along $x$-direction & Total size  & $\Delta y^+$ & Maximum skewness & Maximum aspect ratio\\
\midrule
M1 & 80  & 122,640    & 4.50 & 0.59  & 7.46   \\
M2 & 120 & 430,080    & 3.60 & 0.61  & 7.86   \\
M3 & 180 & 1,451,520  & 2.35 & 0.62  & 7.83   \\
M4 & 270 & 4,898,880  & 1.65 & 0.63  & 7.81   \\
\botrule
\end{tabular}
\end{table*}
\subsection{Mesh independence study of fully-developed turbulent pipe flows}
\label{Sec: MeshIndependence}
LES-WALE simulations for pipe flows are perfromed at $Re_\tau=361$ used four different meshes, named M1, M2, M3, and M4, with cell counts varying from 0.13 to 5 million. Table \ref{table: Mesh information} provides details about these meshes, such as total cell count, cells along the flow direction, maximum skewness and aspect ratio, and $\Delta y^+$ ranges.

 \begin{figure*}[!ht]
\begin{center}
\subfloat[M1]{
\includegraphics[width=0.5\linewidth]{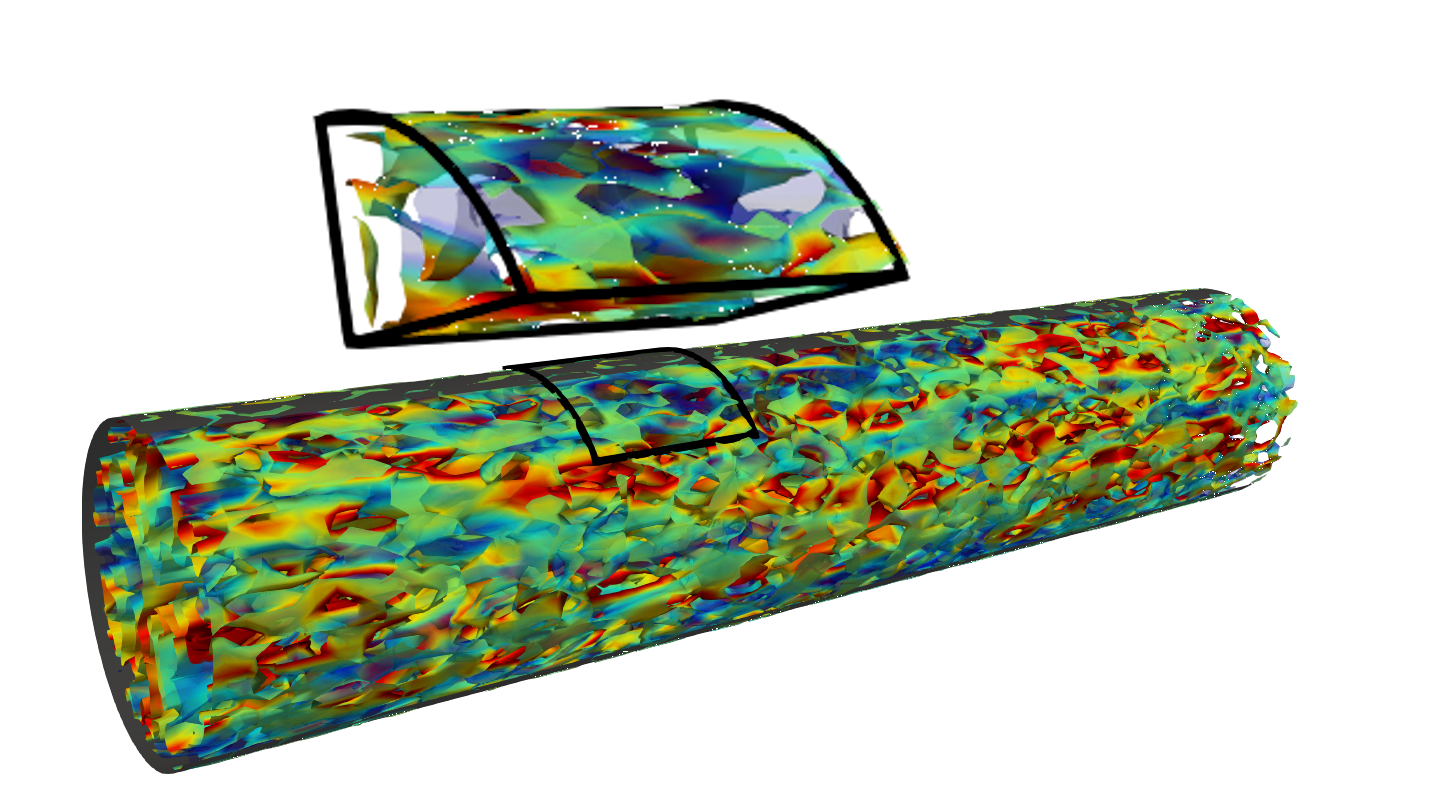}}
\subfloat[M2]{
\includegraphics[width=0.5\linewidth]{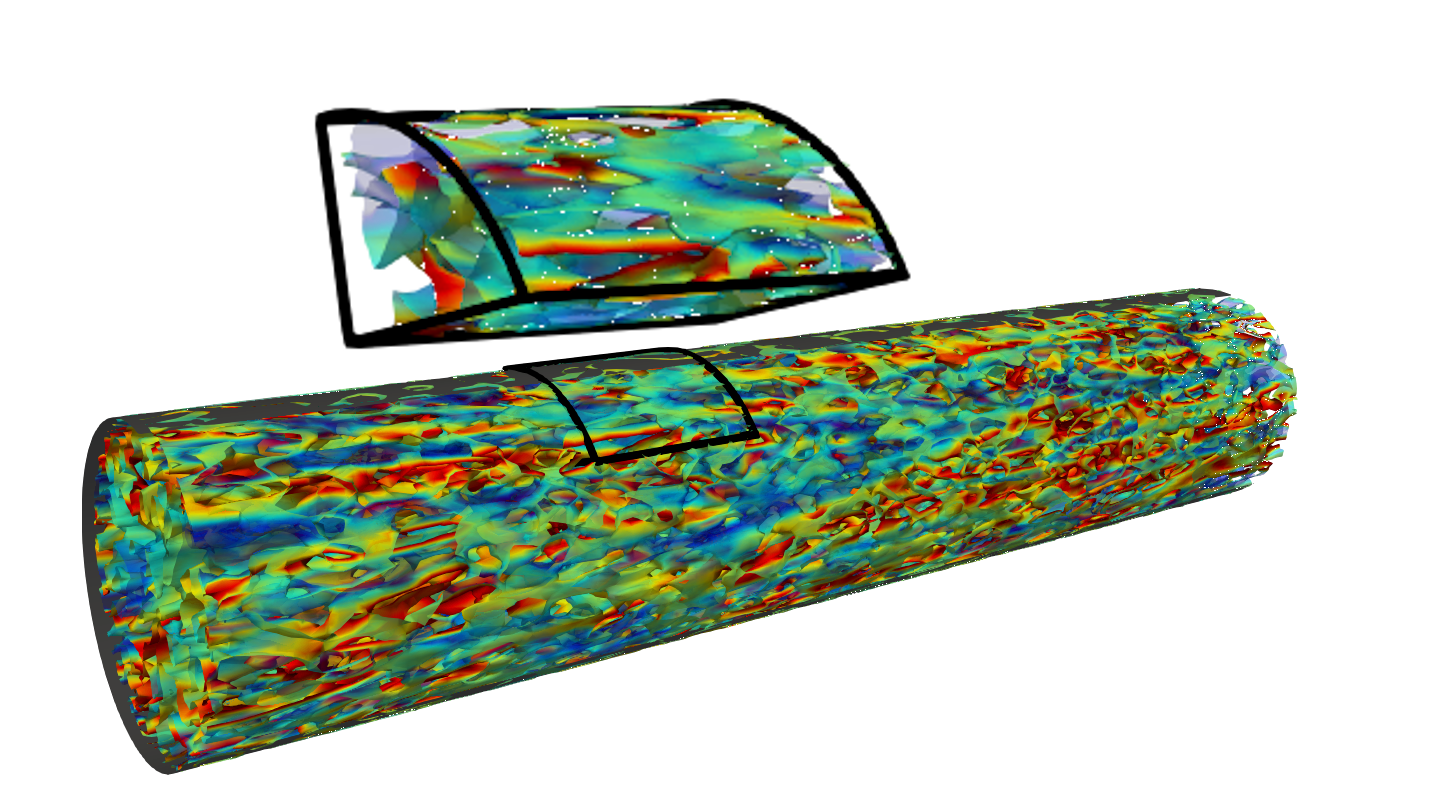}}

\subfloat[M3]{
\includegraphics[width=0.5\linewidth]{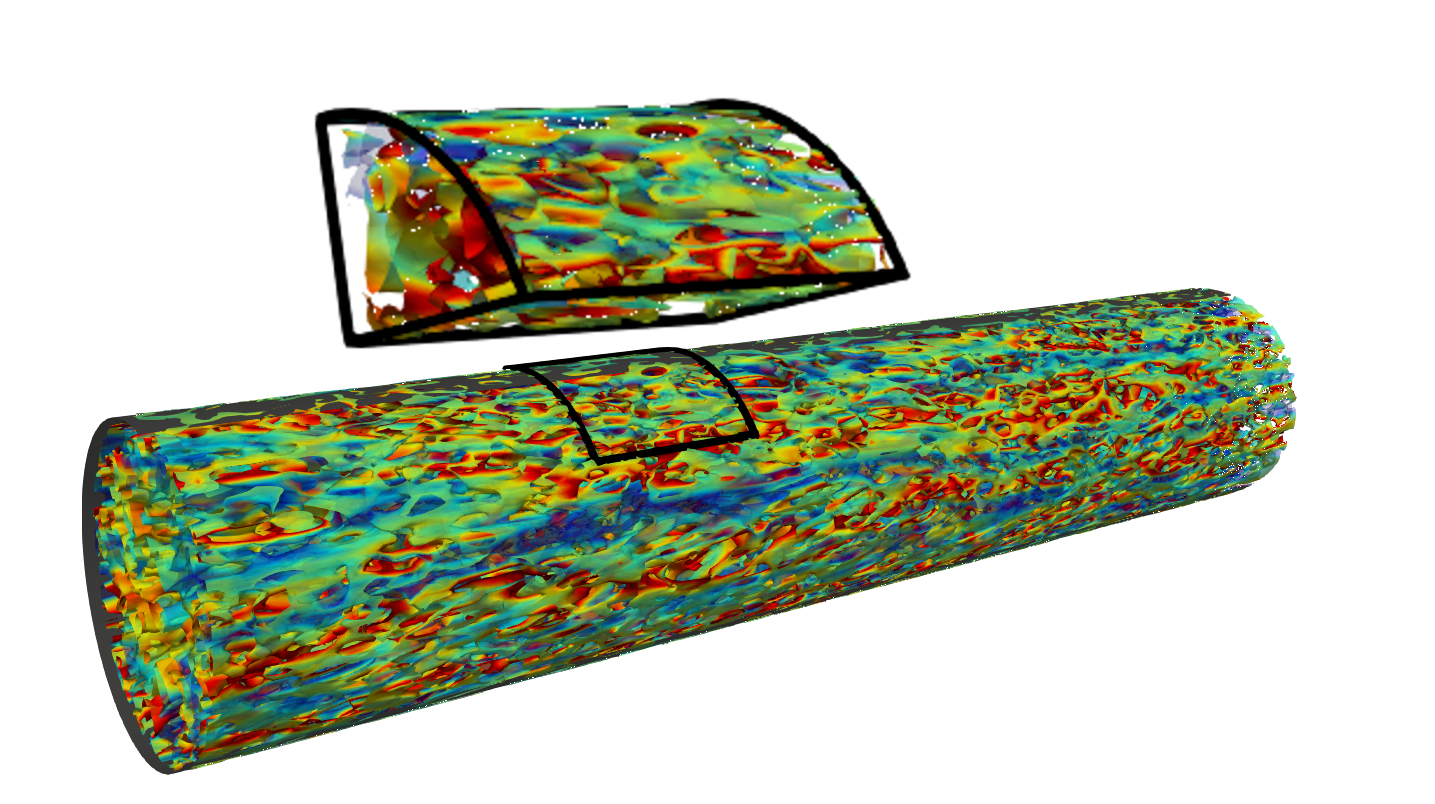}}
\subfloat[M4]{
\includegraphics[width=0.5\linewidth]{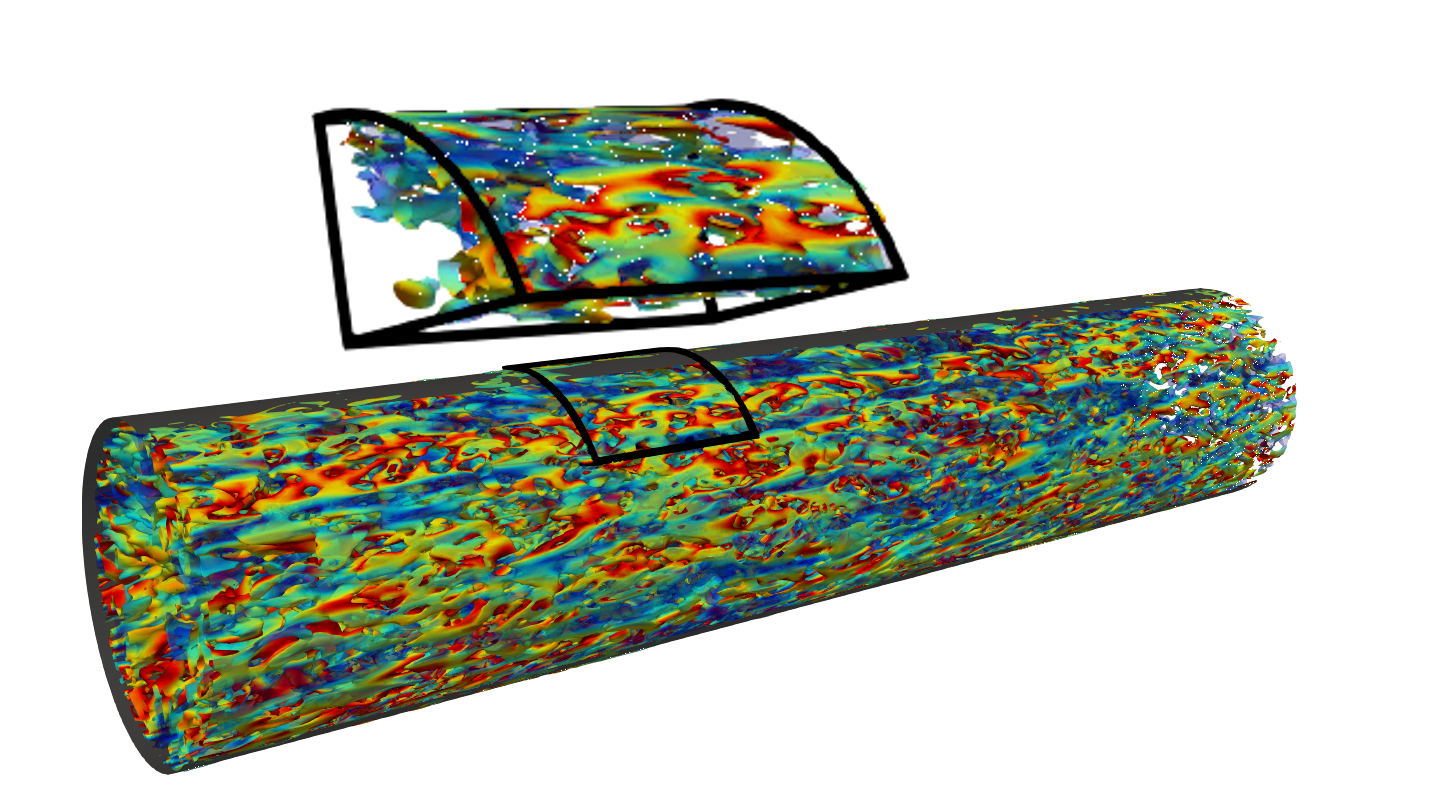}}
\caption{Visualization of vortical structures using the $\lambda_2$-criterion, colored with streamwise velocity fluctuations, $u'$, at $Re_\tau=361$ for (a) M1, (b) M2, (c) M3, and (d) M4. {Color scale ranges from -$0.3$ (blue) to $0.3$ (red) and zoomed-in views are provided for clarity.}}
\label{fig: 0 full view of pipes with vortices and volume rendering for different Meshes}
\end{center}
\end{figure*}

Figure \ref{fig: 0 full view of pipes with vortices and volume rendering for different Meshes} shows the resulting flow using isosurfaces of the second invariant of the velocity gradient tensor, $\lambda_2$ \citep{jeong1995identification}, obtained through the LES-WALE simulations at different mesh resolutions.
Vortices are identified using the $\lambda_2$-criterion and presented for $\lambda_2/\lambda_{2_{\text{max}}} = -0.01$, where more negative values indicate stronger vortices. The pipe shows a dense collection of turbulent eddies from the wall up to $y/R \simeq 0.5$. Coarser grids (M1 and M2) exhibit fewer vortical structures in the boundary layers, forming elongated axial vortices with length scales around $0.1R$. Higher mesh resolution helps capture finer structures near the wall, reducing numerical dissipation. Despite differences in mesh, fluctuations appear similar from M1 to M4. Meshes M3 and M4 show similar turbulence statistics, suggesting convergence and independence from mesh sensitivity.
 
\begin{figure*}[!htp]
\begin{center}
\subfloat[]{
\includegraphics[width=0.45\linewidth]{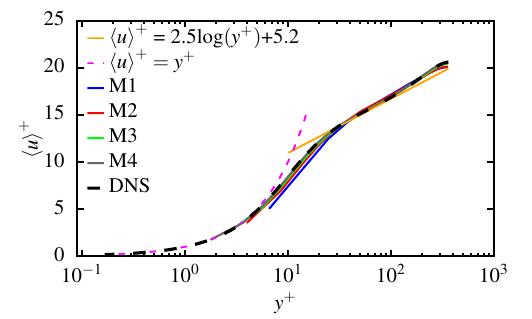}}
\subfloat[]{
\includegraphics[width=0.45\linewidth]{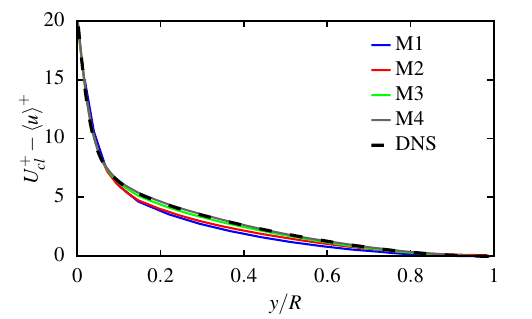}}
\caption{Comparison of LES-WALE and DNS data for mean velocity profiles (a) in inner variables and (b) in velocity-defect form. $U_{cl}^+$ is the normalized centerline velocity. DNS data at $Re_\tau=361$ is shown in dashed black line. The dashed-dotted pink and solid orange lines {in Fig.~\ref{fig: 5 Re361 Mean Flow Profiles grid independence} (a)} represent the universal behavior of the turbulent velocity profile.}
\label{fig: 5 Re361 Mean Flow Profiles grid independence}
\end{center}
\end{figure*}
In Fig. \ref{fig: 5 Re361 Mean Flow Profiles grid independence}, (a) shows mean velocity profiles in inner variables for four grid resolutions and includes DNS data from El Khoury \textit{et al.} \citep{el2013direct} in a pipe flow. The profiles are computed along a vertical line through the pipe center. Higher mesh resolution causes a downward shift in the overlap region compared to DNS data. Although there is a slight difference in the inner layer for M1, the profiles maintain a similar shape across different mesh resolutions. Unlike turbulent channel flows in experiments or numerical simulations \citep{kim1987turbulence,moin1982numerical,bose2010grid,lee2015direct}, the mean velocity profiles do not align with the logarithmic velocity distribution, $\left<u\right>^+ = \left(1/{\kappa}\right)\text{log}(y^+)+B^+$ for $y^+ > 200$, where $\kappa = 0.4$ is the Von K\'arm\'an constant and $B^+ = 5.2$. This discrepancy is likely due to a wake region near the pipe centerline \citep{el2013direct}. Additionally, (b) tests the law of the wall by plotting mean profiles in velocity-deficit form in Fig. \ref{fig: 5 Re361 Mean Flow Profiles grid independence}. Differences in mesh resolution become more evident. Transitioning from M1 to M4, profiles shift toward DNS data, and excellent overall agreement with DNS is achieved for M3 and M4.

\begin{figure*}[!ht]
\begin{center}
\subfloat[]{
\includegraphics[width=0.45\linewidth]{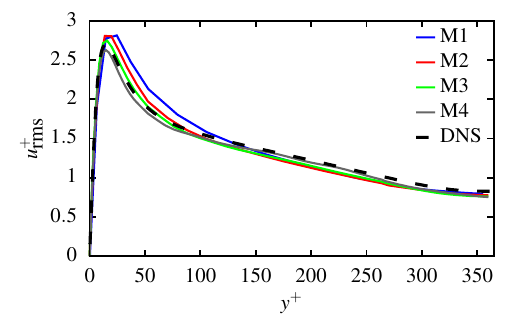}}
\subfloat[]{
\includegraphics[width=0.45\linewidth]{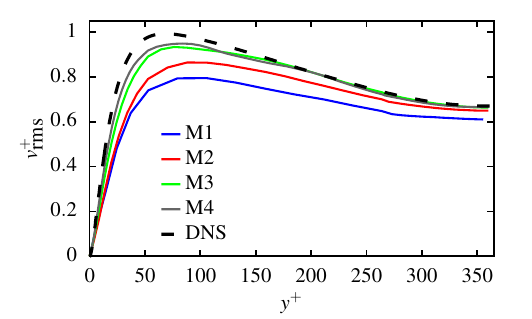}}

\subfloat[]{
\includegraphics[width=0.45\linewidth]{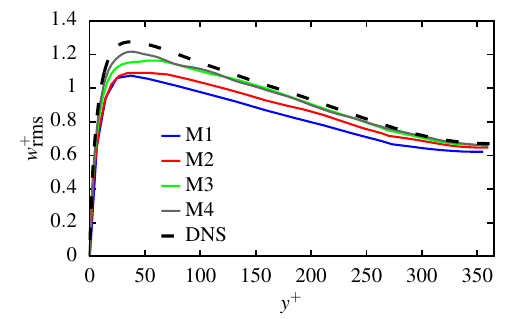}}
\subfloat[]{
\includegraphics[width=0.45\linewidth]{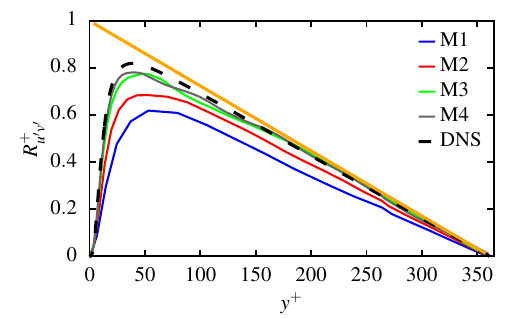}}
\caption{Comparison of LES-WALE and DNS data for (a) axial, $u^+_{\text{rms}}$, (b) wall-normal, $v^+_{\text{rms}}$, and (c) azimuthal, $w^+_{\text{rms}}$, turbulence intensities, and (d) turbulent shear stresses, $R^+_{u'v'}$. The DNS data  at $Re_\tau=361$ are shown in dashed black line. {The thin diagonal line represents the total shear stress, with the difference between this line and the data indicating the viscous shear stress attributed to the mean flow.}}
\label{fig: 5 Re361 Mean fluctuation Profiles grid independence}
\end{center}
\end{figure*}
%
Figure \ref{fig: 5 Re361 Mean fluctuation Profiles grid independence} assesses mesh sensitivity by presenting streamwise, wall-normal, and azimuthal root-mean-squared velocity profiles against the wall distance with inner scaling. As mesh resolution increases, results progressively align with DNS data (dashed black line). Notably, M$1$ and M$2$ show different flow behavior in the outer layer ($y^+ > 50$) compared to M$3$ and M$4$. The finest grid yields excellent agreement with DNS across the entire region, while M$3$ and M$4$ produce similar results despite their resolution difference, indicating statistical independence for these two grids. The positive impact of grid refinement is evident in the gradual increase in result accuracy. Additionally, root-mean-squared velocity fluctuations, $u^+_{\text{rms}}$, $v^+_{\text{rms}}$, and $w^+_{\text{rms}}$, converge toward DNS results with grid refinement. The peak value of $u^+_{\text{rms}}$ consistently occurs at $y^+ \simeq 15$, observed in various studies across wall-bounded flows. In Fig. \ref{fig: 5 Re361 Mean fluctuation Profiles grid independence} (d), profiles of Reynolds shear stress, crucial for momentum transport in wall-bounded turbulence, are presented. Reynolds shear stress near the wall is scaled by classic inner scaling as $u_\tau^2$~\citep{de2000reynolds}. Regardless of mesh resolution, all profiles collapse well for $y^+ \leq 16$ and scale as $u_\tau^2$. As mesh resolution increases, results increasingly agree with DNS data (dashed black line). M3 and M4 stand out, providing exceptionally good results, while M1, being a very coarse mesh, leads to the underprediction of most turbulent properties.

To conclude the mesh sensitivity study, friction factor ($f_D$) values are calculated using $f_D = 8\tau_w/(\rho\left<u\right>^2)$ \citep{chaudhry1979applied} for each simulation and compared with values from the Colebrook correlation~\citep{moody1944friction} in Table \ref{table: friction factor grid dependence}. In all cases, $f_D$ values are slightly underpredicted compared to empirical formulations, as expected due to the nature of turbulence models used in LES, which do not fully resolve wall dynamics. However, with a gradual increase in grid resolution and refinement near the wall, the results improve noticeably. The relative error in the friction coefficient, $f_{D_ {\text{relErr}}} = |f_{D_{\text{num}}}-f_{D_{\text{exp}}}|/f_{D_{\text{exp}}}\times 100\,\%$, tends towards zero. Once again, the measurement of $f_D$ shows that M3 and M4 exhibit similar behavior and outperform other grids in terms of accuracy.       
\begin{table*}[!ht]
\caption{Computed global flow parameters for different grid resolutions. $Re_\tau = u_\tau^0 D/(2\nu)$, and $Re^{ap}_\tau = u_\tau D/(2\nu)$ are the a priori and a posteriori estimates of $Re_\tau$, respectively. $u_\tau^0$ and $u_\tau$ are the a priori and a posteriori estimates of the friction velocity, respectively. $f_{D_\text{num}}$ and $f_{D_{\text{exp}}}$ are numerical estimation and target value of the Darcy friction coefficient, respectively. $f_{D_{\text{relErr}}}$ the relative error in the numerical computation of $f_D$.}\label{table: friction factor grid dependence}
\begin{tabular}{p{2cm} p{2cm} p{2cm} p{2cm} p{2cm} p{2cm} p{2cm}}
\hline
Mesh & $Re_b$ & $Re_\tau$ & $Re^{ap}_\tau$ & ${f_D}_{\text{num}}$ & ${f_D}_{\text{exp}}$ & ${f_D}_{\text{relErr}}$ \\ \hline
M1 &  11,700  &  361  &  322  &  0.0242  &  0.0296  & 18\%  \\[0.2cm]
M2 &  11,700  &  361  &  343  &  0.0279  &  0.0296  & 6\%   \\[0.2cm]
M3 &  11,700  &  361  &  357  &  0.0298  &  0.0296  & 1\%   \\[0.2cm]
M4 &  11,700  &  361  &  360  &  0.0303  &  0.0296  & 2.4\%   \\[0.2cm]
DNS &  11,700  &  361  &  361  &  0.03040  &  0.0296  & 2.7\%   \\
\hline
\end{tabular}
\end{table*} 
\subsection{Different subgrid model behavior for ${Re_\tau=361}$}
\label{Sec: SGS model dependence}
To evaluate the predictive capabilities of the subgrid models, numerical simulations of a turbulent pipe flow at a fixed turbulent $Re_\tau=361$, are performed, comparing results against previously reported DNS findings \citep{el2013direct}. This comparison involves five different subgrid models (SMG, OEEVM, WALE, LDKM, and DSEM). 
{As we refine the mesh resolution, the results from the LES gradually approach the DNS results, as shown in Figs.~\ref{fig: 5 Re361 Mean Flow Profiles grid independence} and \ref{fig: 5 Re361 Mean fluctuation Profiles grid independence} of Sec. \ref{Sec: MeshIndependence}. The reason for this is that as $\Delta$ decreases $k_{sgs}$ and $\nu_{sgs}$ also decrease approximately as $\Delta^2$. 
Since the objective of this sensitivity study is to examine the influence of the subgrid model, it is advisable to use the coarser meshes. The coarser meshes will emphasize the differences between the subgrid models more effectively. On finer meshes, we will not be able to detect any noticeable differences between the models. Therefore, to facilitate an effective comparison, mesh M2 has been selected.}

\begin{table*}[!ht]
\caption{Computed global flow parameters for different subgrid models. $Re_\tau = u_\tau^0 D/(2\nu)$, and $Re^{ap}_\tau = u_\tau D/(2\nu)$ are the a priori and a posteriori estimates of $Re_\tau$, respectively. $u_\tau^0$ and $u_\tau$ are the a priori and a posteriori estimates of the friction velocity, respectively. $f_{D_\text{num}}$ and $f_{D_{\text{exp}}}$ are numerical estimation and target value of the Darcy friction coefficient, respectively. $f_{D_{\text{relErr}}}$ the relative error in the numerical computation of $f_D$.}\label{table: friction factor subgrid dependence}
\begin{tabular}{p{2cm} p{2cm} p{2cm} p{2cm} p{2cm} p{2cm} p{2cm} p{2cm}}
\hline
Subgrid model & Mesh & $Re_b$ & $Re_\tau$ & $Re^{ap}_\tau$ & ${f_D}_{\text{num}}$ & ${f_D}_{\text{exp}}$ & ${f_D}_{\text{relErr}}$ \\\hline
OEEVM   & M2 & 11,700  &  361  &  344  &  0.0276  &  0.0296  & 7\% \\[0.2cm]
SMG     & M2 & 11,700  &  361  &  342  &  0.0274  &  0.0296  &  7\% \\[0.2cm]
WALE    & M2 & 11,700  &  361  &  345  &  0.0279  &  0.0296  &  6\% \\[0.2cm]
LDKM    & M2 & 11,700  &  361  &  345  &  0.0278  &  0.0296  &  6\% \\[0.2cm]
DSEM    & M2 & 11,700  &  361  &  374  &  0.0327  &  0.0296  &  10\%  \\[0.2cm]
DNS &  &11,700  &  361  &  361  &  0.03040  &  0.0296  & 2.7\%   \\
\hline
\end{tabular}
\end{table*}
Table \ref{table: friction factor subgrid dependence} presents a priori and a posteriori estimates of $Re_\tau$, estimates of $f_{D_{\text{num}}}$, $f_{D_{\text{exp}}}$, and $f_{D_{\text{relErr}}}$ for each subgrid model. It is noted that, across all cases, the simulated $f_D$ values are slightly underpredicted compared to targeted values. However, no discernible trend linking $f_D$ with subgrid models is evident, with negligible differences in $f_D$ values among the various models. Although additional refinement near the wall could potentially enhance results, the impact on different models remains uncertain. Nevertheless, in line with the LES philosophy, such \emph{improvement} would come at the expense of increased computational requirements -- an outcome undesirable in the context of this subgrid model comparison.
 
\begin{figure*}[!htp]
\begin{center}
\subfloat[]{
\includegraphics[width=0.45\linewidth]{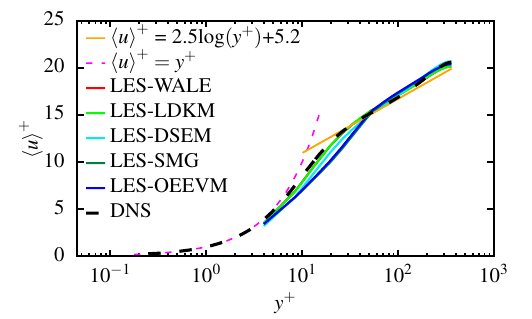}
\label{fig: 6 Re361 Mean Flow Profiles subgrid models - 1}}
\subfloat[]{
\includegraphics[width=0.45\linewidth]{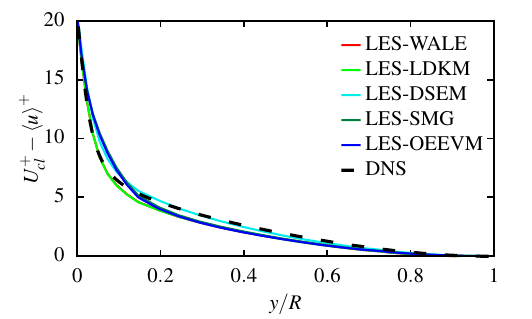}
\label{fig: 6 Re361 Mean Flow Profiles subgrid models - 2}}
\caption{(a) Mean velocity profiles  and (b) mean velocity-defect profiles for $Re_\tau=361$ and five different subgrid models, normalized by ${u^0_\tau}$ which is a priori estimate of the friction velocity. $U_{cl}^+$ is the normalized centerline velocity. The dashed black line presents DNS data from El Khoury \textit{et al.} \citep{el2013direct} and the dashed pink and solid orange lines {in Fig.~\ref{fig: 6 Re361 Mean Flow Profiles subgrid models} (a)} represent the universal behavior of the turbulent velocity profile.}
\label{fig: 6 Re361 Mean Flow Profiles subgrid models}
\end{center}
\end{figure*}
To assess predictive abilities in both near-wall and outer regions, we calculate mean velocity and mean velocity-defect profiles normalized in wall units. Fig. \ref{fig: 6 Re361 Mean Flow Profiles subgrid models} (a) displays mean velocity profiles in wall units, while Fig. \ref{fig: 6 Re361 Mean Flow Profiles subgrid models} (b) shows mean velocity-defect profiles in the outer region for different subgrid models. Analyzing Fig. \ref{fig: 6 Re361 Mean Flow Profiles subgrid models}, all subgrid models exhibit distinct behaviors, clustering based on their predictive capabilities. Near the wall, LES-WALE and LES-LDKM accurately predict mean velocity profiles, while LES-DSEM is also accurate but slightly less so. In contrast, LES-SMG and LES-OEEVM slightly underpredict near-wall mean flow but show reasonable agreement with DNS data and other models in the core flow region.

Our computational models yield slightly lower $Re_\tau$ than El Khoury \textit{et al.} \citep{el2013direct}. This results in slight overestimates in mean velocity profiles when normalized using inner units compared to DNS data \citep{el2013direct}. In Fig. \ref{fig: 6 Re361 Mean Flow Profiles subgrid models} (b), showing mean velocity-defect profiles, there is a slight underestimation compared to DNS values. This occurs as we used the theoretical value of {$u_\tau$ as a normalization parameter instead of the actual value measured in each simulation, see Fig. \ref{fig: 4b_rebuttal} in Appendix \ref{Appendix 0}}. For $y/R > 0.65$, all curves collapse onto DNS results. Additionally, all mean velocity profiles exhibit a log-law behavior above the buffer layer, i.e., $y^+ \geq 30$ (see the orange solid line in Fig. \ref{fig: 6 Re361 Mean Flow Profiles subgrid models} (a)). In our results, the logarithmic representation of turbulent profiles shows a universal behavior across subgrid models, with values $1/{\kappa}=2.5$ and $B^+=5.2$, proposed and accepted for various canonical flows. 
\begin{figure*}[!htp]
\begin{center}
\subfloat[]{
\includegraphics[width=0.45\linewidth]{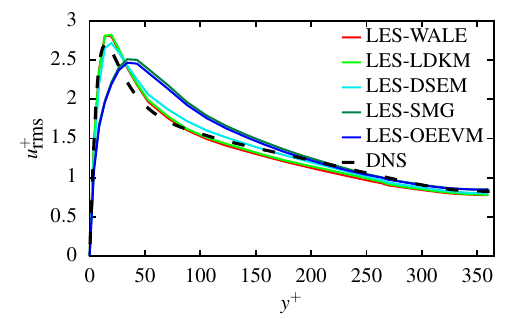}
\label{fig: 6a Re361 Mean Flow Profiles subgrid models - 1}}
\subfloat[]{
\includegraphics[width=0.45\linewidth]{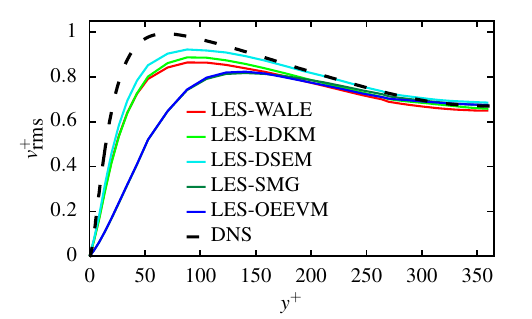}
\label{fig: 6a Re361 Mean Flow Profiles subgrid models - 2}}

\subfloat[]{
\includegraphics[width=0.45\linewidth]{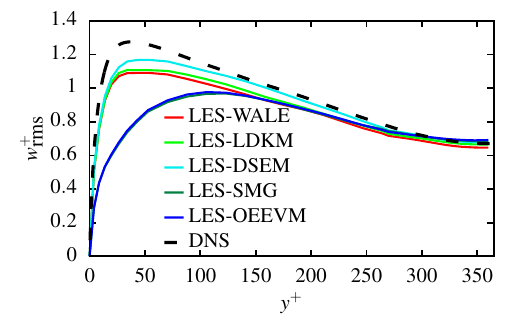}
\label{fig: 6a Re361 Mean Flow Profiles subgrid models - 3}}
\subfloat[]{
\includegraphics[width=0.45\linewidth]{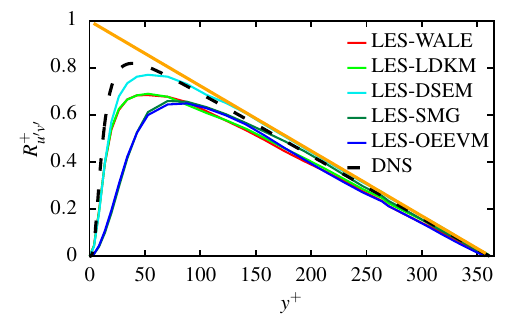}
\label{fig: 6a Re361 Mean Flow Profiles subgrid models - 5}}
\caption{Comparison of (a) axial, $u^+_{\text{rms}}$, (b) wall-normal, $v^+_{\text{rms}}$, and (c) azimuthal, $w^+_{\text{rms}}$, turbulence intensities; 
(d) turbulent Reynolds shear stress, $R^+_{u'v'}$, with DNS data (dashed lines) as functions of $y^+$, at $Re_\tau=361$, for different subgrid models. {The thin diagonal line represents the total shear stress, with the difference between this line and the data indicating the viscous shear stress attributed to the mean flow.}}
\label{fig: 6a Re361 Mean Flow Profiles subgrid models}
\end{center}
\end{figure*}
RMS velocity fluctuation values are shown in Figs. \ref{fig: 6a Re361 Mean Flow Profiles subgrid models} (a-c). LES-SMG and LES-OEEVM underpredict peak values and overpredict their locations compared to LES-WALE, LES-LDKM, LES-DSEM, and DNS. LES-WALE, LES-LDKM, and LES-DSEM accurately predict peak locations but overestimate the values. This aligns with observations in other channel flow studies where $u^+_{\text{rms}}$ tends to be overpredicted in LES compared to DNS \citep{bose2010grid}. For $v^+_{\text{rms}}$ and $w^+_{\text{rms}}$, all subgrid models exhibit excellent qualitative behavior compared to DNS, as shown in Figs. \ref{fig: 6a Re361 Mean Flow Profiles subgrid models} (b-c). However, quantitatively, all subgrid models underpredict the fluctuation values. LES-OEEVM and LES-SMG are highly diffusive near the wall and fail to predict peak values accurately. In contrast, LES-LDKM and LES-WALE behave closely to DNS but with some underprediction. Surprisingly, LES-DSEM emerges as the most accurate candidate when compared to DNS. Both profiles of velocity fluctuations (cyan data in Figs. \ref{fig: 6a Re361 Mean Flow Profiles subgrid models} (b-c)) show good agreement with DNS due to their anisotropic nature, albeit with slight underprediction. This may be related to fixed values of coefficients $c_m$ and $c_e$ in the kinetic energy distribution equation (Eq. \eqref{eq:11}) in the DSEM subgrid model. While dynamically computing these coefficients could improve predictions, such enhancements are beyond the scope of this paper.

$R^+_{u'v'}$ profiles in Fig. \ref{fig: 6a Re361 Mean Flow Profiles subgrid models} (d) show similar behavior to DNS for LES-WALE, LES-LDKM, and LES-DSEM models. In the viscous wall region ($y^+ < 15$) and outer region ($y^+> 50$), $R^+_{u'v'}$ becomes linear. Quantitatively, LES-WALE and LES-LDKM reproduce DNS tendencies, including a peak at $y^+ \simeq 30$ and semi-linear behavior in the outer region. Both subgrid models accurately predict the peak location, albeit with a slight underestimation of its value. Surprisingly, LES-DSEM closely agrees with DNS compared to LES-WALE and LES-LDKM. In contrast, LES-SMG and LES-OEEVM tend to overpredict the peak location while underpredicting the peak value, consistent with rms values of velocity fluctuations in Figs. \ref{fig: 6a Re361 Mean Flow Profiles subgrid models} (a-c).
\subsection{Investigation of fully-developed turbulent flows for $Re_\tau$ ranging from $361$ to $2, 000$}\label{sec:VTM}
\label{Sec: Re dependence}
The analysis thus far suggests that the DSEM, WALE, and LDKM subgrid models accurately predict flow behavior. Therefore, we proceed with the WALE model for further exploration at high $Re_\tau$ values. We choose not to use LDKM and DSEM, as our earlier observation suggests similar prediction capabilities among WALE, LDKM, and DSEM, simplifying the computational process. This simplifies the computational process by selecting one model. Results are compared against previously reported DNS results \citep{el2013direct,pirozzoli2021one} for four different $Re_\tau$ values: $361$, $550$, $1,000$, and $2,000$. In all simulations, $Re_\tau$ values are initially undefined and computed a posteriori, with $Re_b$ values taken from El Khoury \textit{et al.} \citep{el2013direct} and Pirozzoli \textit{et al.} \citep{pirozzoli2021one}. To explore near-wall-modeled LES behavior under coarse grid conditions and high $Re_\tau$, we choose the coarse M2 grid as the base grid for $Re_\tau=361$, where approximately 85\% of turbulent Reynolds shear stresses and turbulent kinetic energy are resolved. Following Pope \citep{Pope_2004}, this implies using a grid marginally acceptable for LES in the high $Re$ study, expecting errors to be at least as high as for $Re_\tau=361$ or possibly higher. 
\subsubsection{Mesh refinement based on Taylor scaling}
To resolve a relatively similar amount of turbulent Reynolds shear stresses and turbulent kinetic energy for higher $Re_\tau$, we estimate the integral, Taylor, and Kolmogorov scales using conventional approximations. 
The integral length scale, $l_I$, is set as the pipe diameter, i.e., $l_I=D$. 
The Taylor scale, $\lambda$, an intermediate length scale in the inertial subrange, is estimated as $\lambda = \sqrt{(15 \nu/\epsilon)}\mathfrak{u}_{rms}$ \citep{taylor1935statistical}. Here, $\mathfrak{u}_{rms} = \sqrt{\left(u'^2+v'^2+w'^2\right)/3}$ is the root mean square of the velocity fluctuations. The Kolmogorov scale, $\eta$, dissipation subrange, is estimated as $\eta = (\nu^3/\epsilon)^{1/4}$ \citep{pope2000turbulent}. Combining the expression for the Kolmogorov scale with the dissipation rate, $\epsilon \simeq k^{3/2}/l_I$, where $k=\frac{3}{2}{\mathfrak{u}}_{rms}^2$, yields $\eta/l_I = C(\mathfrak{u}_{rms}l_I/\nu)^{-3/4} = C Re_l^{-3/4}$, with $C  \simeq 0.86$ as a constant and $Re_l$ as the integral scale $Re$ number. The value of $\mathfrak{u}_{rms}$, can be approximated using the turbulent intensity, $I$, as $\mathfrak{u}_{rms}=I U_b$, where $I=10\%$ for high turbulence intensity applications. The grid resolution needed to resolve a relatively similar amount of the turbulent Reynolds shear stresses and the turbulent kinetic energy as at $Re_\tau=361$ for higher $Re_\tau$ values is computed from the expression of $\lambda/l_I$ and provided in Table \ref{table: friction factor reynolds dependence}.
\begin{table*}[!ht]
\caption{Computed global flow parameters for different turbulent $Re_\tau$. $Re_\tau = u_\tau^0 D/(2\nu)$, and $Re^{ap}_\tau = u_\tau D/(2\nu)$ are the a priori and a posteriori estimates of $Re_\tau$, respectively. $u_\tau^0$ and $u_\tau$ are the a priori and a posteriori estimates of the friction velocity, respectively. $f_{D_\text{num}}$ and $f_{D_{\text{exp}}}$ are numerical estimation and target value of the Darcy friction coefficient, respectively. $f_{D_{\text{relErr}}}$ the relative error in the numerical computation of $f_D$.}\label{table: friction factor reynolds dependence}
\begin{tabular}{p{2cm} p{2cm} p{2cm} p{2cm} p{2cm} p{2cm} p{2cm} p{2cm} p{2cm} p{2cm}}
\hline
 Grids & $Re_b$ & Mesh total size & $Re_\tau$ & $Re^{ap}_\tau$ & ${f_D}_{\text{num}}$ & ${f_D}_{\text{exp}}$ & ${f_D}_{\text{relErr}}$ \\
\hline
G1 & 11,700   &   0.5M cells   &  361   &  345    &  0.0279  &  0.0296  &  5\% \\[0.2cm]
G2 & 19,000   &   1.0M cells   &  550   &  517    &  0.0237  &  0.0262  &  9\% \\[0.2cm]
G3 & 37,700   &   2.3M cells   &  1,000 &  928    &  0.0194  &  0.0223  &  13\% \\[0.2cm]
G4 & 82,500   &   6.5M cells   &  2,000 &  1,780  &  0.0149  &  0.0187  &  20\% \\[0.2cm]
\hline
\end{tabular}
\end{table*}  

Table \ref{table: friction factor reynolds dependence} presents a priori and a posteriori $Re_\tau$ estimations. The chosen $Re_b$ value is deemed adequate. As expected, $Re_\tau$ is underpredicted in all simulations (Table \ref{table: friction factor reynolds dependence}). However, the earlier grid resolution analysis suggests significant improvement with refinement, indicating eventual convergence to $Re_\tau$. Numerical estimations of $f_{D_{\text{num}}}$ for various turbulent $Re_\tau$, along with $f_{D_{\text{exp}}}$ values, are in Table \ref{table: friction factor reynolds dependence}. Deviations of approximately 5-20\% from expected results are observed.

\begin{figure*}[!htp]
\begin{center}
\subfloat[]{
\includegraphics[width=0.45\linewidth]{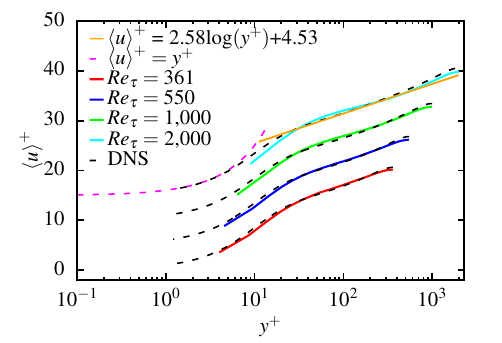}}
\subfloat[]{
\includegraphics[width=0.45\linewidth]{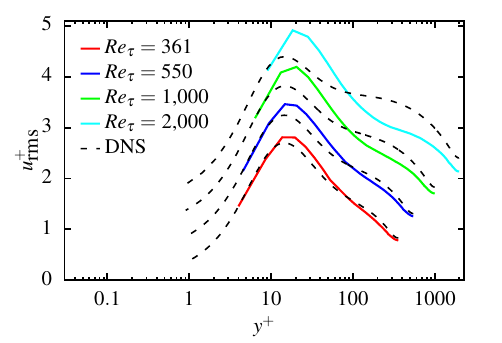}}
\caption{Comparison of (a) $\left<u\right>^+$ and (b) $u_{\text{rms}}^+$, as functions of $y^+$ for different values of $Re_\tau$ and {the WALE subgrid model.} $u_{\text{rms}}^+$ profiles are plotted with an offset of $\Delta \left < . \right >^+ = 1$ for clarity. In all figures, the DNS data are represented in dashed black lines.}
\label{fig: 15 Bad Smooth mean flow profiles for different Reynolds}
\end{center}
\end{figure*}
In Fig. \ref{fig: 15 Bad Smooth mean flow profiles for different Reynolds} (a), mean velocity profiles from LES-WALE are compared to DNS data for various $Re_\tau$ values. A generally good agreement is observed, especially in the inner layer and overlap regions, with data gradually converging to a logarithmic distribution. Fig. \ref{fig: 15 Bad Smooth mean flow profiles for different Reynolds} (b) shows profiles of streamwise turbulent velocity fluctuations normalized by $u_\tau^2$ in relation to $Re_\tau$. For low $Re_\tau$, results match DNS profiles, but as $Re_\tau$ increases, differences between LES-WALE and DNS become more pronounced. Specifically, LES-WALE consistently overpredicts the peak value of $u^+_{\text{rms}}$ for each $Re_\tau$, intensifying with higher $Re_\tau$. Conversely, in the outer region, LES-WALE tends to underestimate velocity fluctuations, more noticeable with increasing $Re_\tau$. Meshes based on Taylor scaling do not capture fluctuations' overshoot in the outer region observed in DNS. Results from Berrouk \textit{et al.}\citep{berrouk2007stochastic} and Ferrer \textit{et al.}\citep{ferrer2019high} also show discrepancies with DNS data. This could be due to our grid resolution and seems inherent in LES subgrid models, as seen in our previous grid resolution study, even for grid M3. A more detailed investigation with advanced subgrid models like DSEM is recommended, although beyond the scope of this study.
\subsubsection{Optimizing Mesh: Directional Refinement for Improved Results}
To address the mesh resolution issue related to large-scale fluctuations, we revisited our meshing strategy, initially based on Taylor's scale. We conducted simulations at $Re_\tau=1000$ for several new generated grids (G31, G32, G33, G34) with decreasing mesh spacing in all directions. Similar meshes were generated for $Re_\tau=361, 550, 2,000$. Additional details on the grids are available in Table. \ref{table: New Mesh information}. Results for M1-M4 are already presented in Sec. \ref{Sec: MeshIndependence}.
\begin{table*}[!ht]
\caption{Parameters of the new grid used for the grid convergence study; mesh spacings are given in wall units; $\Delta x^+$, $\Delta z^+$: mesh spacings in the streamwise and azimuthal directions; $\Delta y^+_{wall}$, $\Delta y^+_{center}$: mesh spacings in the wall normal direction at the wall and at the center of the channel.}\label{table: New Mesh information}
\begin{tabular}{p{2cm} p{2cm} p{2cm} p{2cm} p{2cm} p{2cm} p{2cm} p{2cm}}
\hline
Grids & $Re_\tau$ & Mesh total size & $\Delta x^+$ & $\Delta y^+_{wall}$ & $\Delta y^+_{center}$ & $\Delta z^+$  \\
\hline
M1    &  361   &   0.13M cells   &  40   &  5    &  8  &  24 \\ 
M2/G1    &  361   &   0.5M cells   &  28  &  3    &  6  &  16 \\
M3    &  361   &   1.5M cells   &  20   &  3    &  4  &  12 \\
M4    &  361   &   5M cells   &  13   &  2    &  3  & 4 \\
      &   &      &     &      &    &  \\
G2    &  550   &   1M cells   & 32   &  4    &  6  & 19 \\
G21   &  550   &   8M cells   & 17   &  2    &  3  & 10 \\
      &        &      &     &      &    &  \\
G3    &  1000   &   2.3M cells   & 44   &  6    &  9  & 26 \\
G31   &  1000   &   2.8M cells   & 42   &  1e-2   &   8 &  24\\
G32   &  1000   &   8M cells   & 44   &  7e-2   &   4 &  11\\
G33   &  1000   &   15M cells   & 28   &  3    &  5  & 15 \\
G34   &  1000   &   36M cells   & 19   &  2    &  4  & 11 \\
      &        &      &     &      &    &  \\
G4   &  2000   &   6.5M cells   & 59   &  7    &  12  & 35 \\
G41   &  2000   &   218M cells   & 19   &  2    &  4  & 10 \\
\hline
\end{tabular}
\end{table*} 
\begin{figure*}[!htp]
\begin{center}
\subfloat[]{
\includegraphics[width=0.45\linewidth]{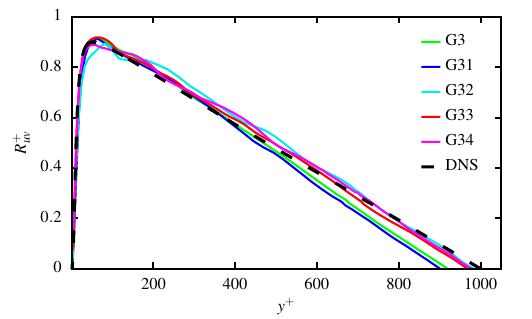}}
\subfloat[]{
\includegraphics[width=0.45\linewidth]{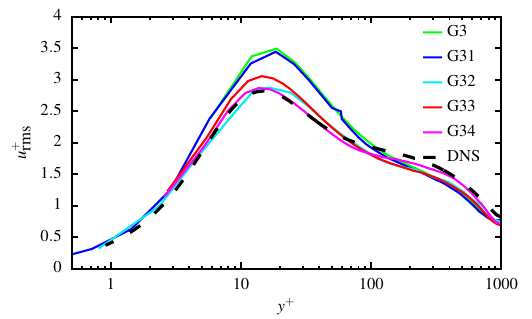}}
\caption{(a) Turbulent shear stresses ($R^+_{u'v'}$) and (b) axial turbulent velocity fluctuations ($u_{\text{rms}}^+$) on a semi-logarithmic scale, are shown as a function of $y^+$ at $Re_\tau=1,000$ {for the WALE subgrid model.} In all subfigures, the DNS data are represented in dashed black lines.}
\label{fig: 23 Smooth mean flow profiles for Retau=1000}
\end{center}
\end{figure*}

Figure \ref{fig: 23 Smooth mean flow profiles for Retau=1000} displays mean turbulent shear stress and rms streamwise velocities for various meshes at $Re_\tau=1000$, along with DNS data for comparison. G32, G33, and G34 simulations show collapsed profiles for turbulent shear stress, while G3 and G31, using coarser resolution based on Taylor's scale, underestimate the results. This suggests grid convergence is achieved for $\Delta y^+_{\text{wall}} \sim 3, \Delta z^+ \sim 15, \Delta x^+ \sim 30$. However, for similar resolution, Fig. \ref{fig: 23 Smooth mean flow profiles for Retau=1000} (b) shows that rms velocity fluctuations are not collapsed. Good agreement with DNS data is only observed with G34. Comparing profiles for G3, G31, G32, and G33, it is observed that to accurately capture small-scale fluctuations, especially overshoots in the outer region, in the presence of wall models, $\Delta y^+$ values can be relaxed, but $\Delta x^+$ and $\Delta z^+$ values play a significant role and must be strictly constrained. A trend can even be established between the peak rms velocity and the grid resolution: the former improves when the mesh resolution is respected as follows: $\Delta z^+ \leq 10, \Delta x^+ = 20-25$. Moreover, the G34 mesh resolution is determined by the Kolmogorov scales, specifically the grid points utilized in LES-WALE, denoted as $N_{LES} \approx (2.25 \cdot Re_\tau)^{9/4}$. It is worth noting that the factor $2.25$ differs from the one employed in DNS studies, which is $3$~\citep{wilcox1998turbulence}.

\begin{figure}[!htp]
\begin{center}
\subfloat[$Re_\tau=361$]{
\includegraphics[width=0.95\linewidth]{1dV3.png}}

\subfloat[$Re_\tau=550$]{
\includegraphics[width=0.95\linewidth]{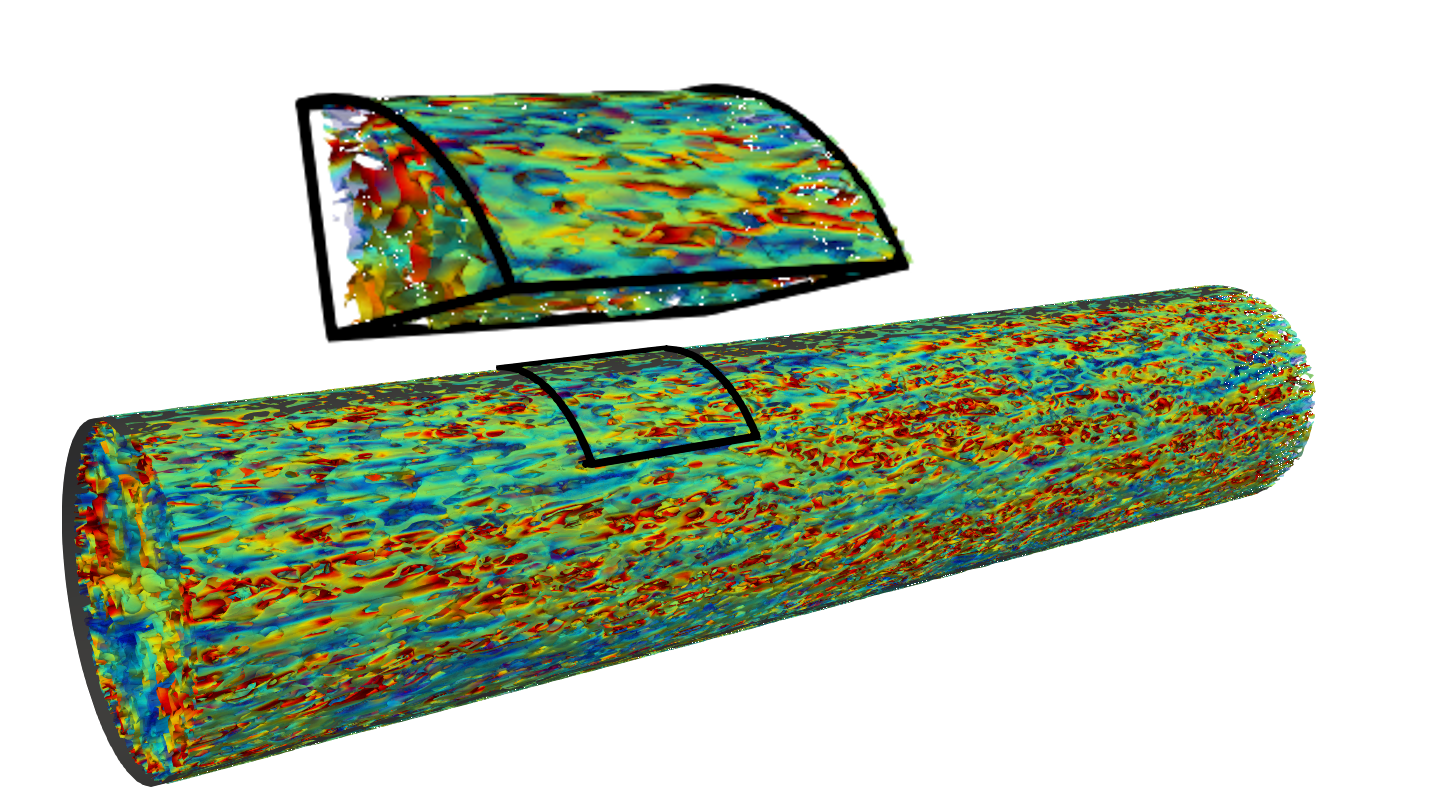}}

\subfloat[$Re_\tau=1,000$]{
\includegraphics[width=0.95\linewidth]{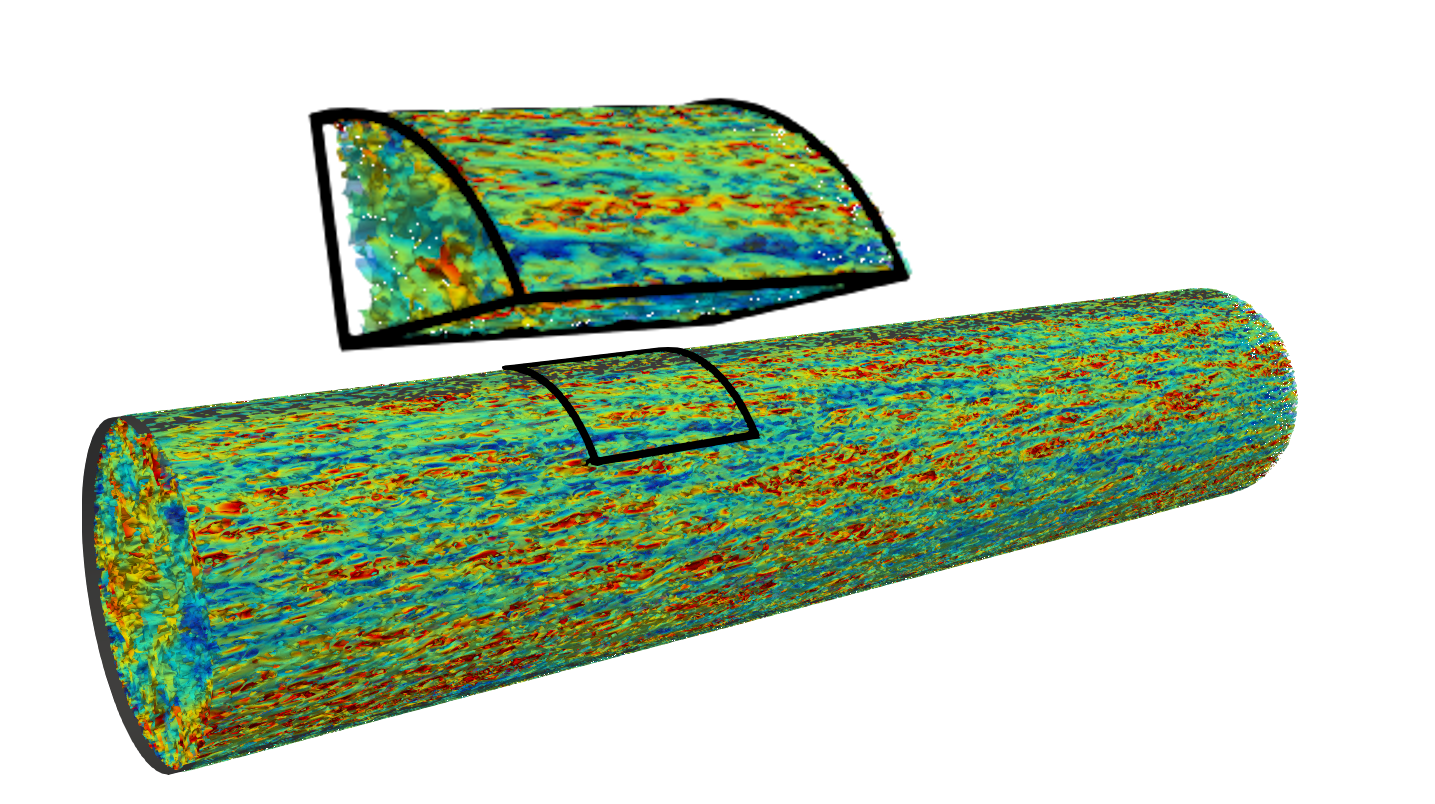}}

\subfloat[$Re_\tau=2,000$]{
\includegraphics[width=0.95\linewidth]{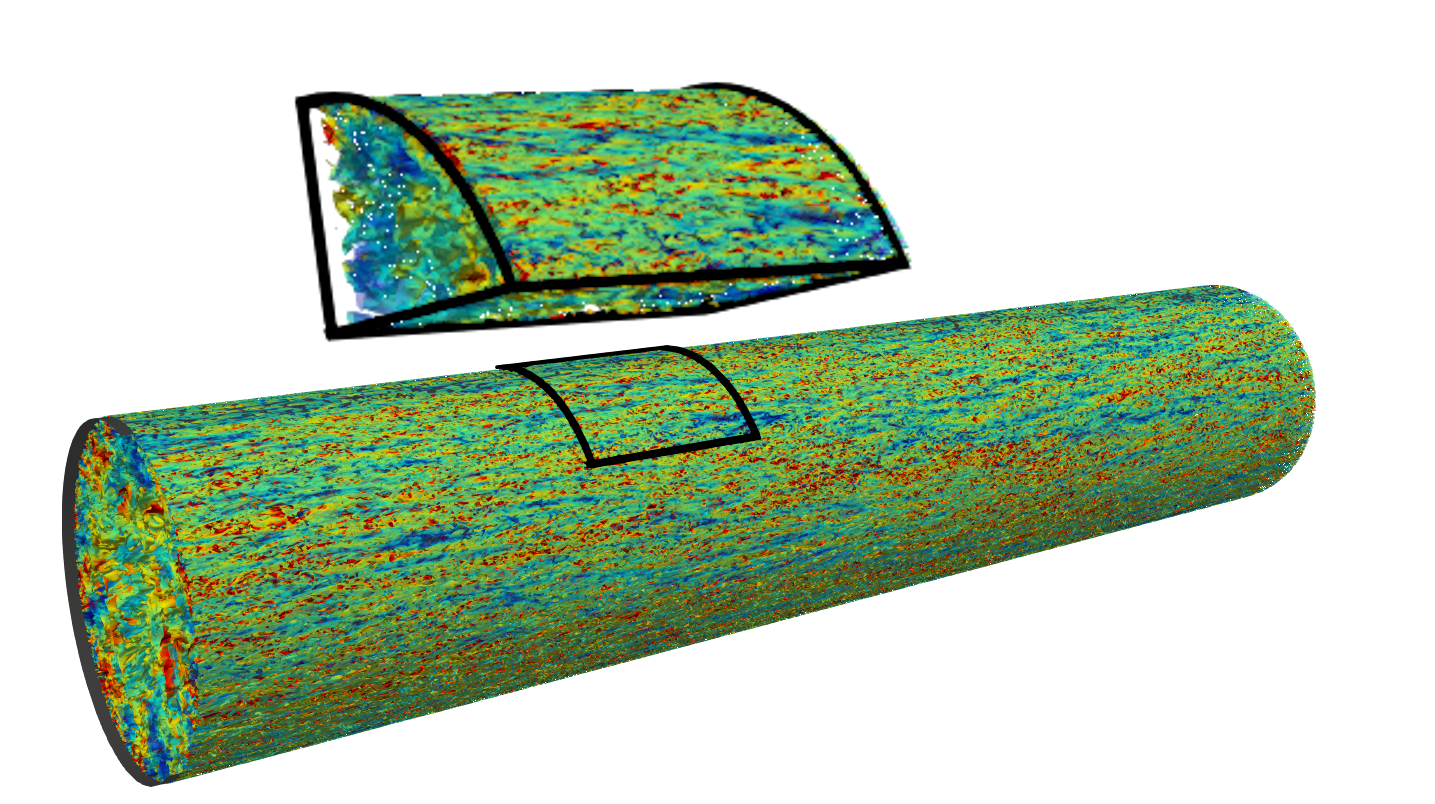}} 
%
\caption{Visualization of vortical structures using the $\lambda_2$-criterion, color-coded by streamwise velocity fluctuations, $u'$ (a) $Re_\tau=361$ (M4 mesh), (b) $Re_\tau =550$ (G21 mesh), (c) $Re_\tau=1,000$ (G34 mesh), and (d) $Re_\tau=2,000$ (G41 mesh). {Color scale ranges from -$0.3$ (blue) to $0.3$ (red) and zoomed-in views are provided for clarity.}}
\label{fig: 14 full view of pipes with vortices and volume rendering for different Reynolds}
\end{center}
\end{figure}

Before doing further quantitative analysis, we show the evolution of the vortical structures for different $Re_\tau$ values in Fig. \ref{fig: 14 full view of pipes with vortices and volume rendering for different Reynolds}. Vortices, identified using the $\lambda_2$-criterion and plotted for $\lambda_2/\lambda_{2_{\text{max}}} = -0.01$, appear densely along the pipe from the wall upto the wall-normal location $y/R \simeq 0.5$. Notably, with increasing $Re_\tau$, fluctuations and resolved vortical structures become finer and more amplified. Streaks become discernible in the near-wall cylindrical shells, arranged in a cross-stream pattern. Regardless of turbulent $Re$, low-speed streaks (deep blue color, approximately the size of the pipe radius) align with large-scale ejections, while high-speed streaks (red-yellow color, of same size as low-speed streaks) correspond to substantial inflow from the core flow. Smaller streaks appear, corresponding to ejection and sweep events in the buffer layer. The flow organization on at least two length scales is apparent, with the separation increasing as $Re_\tau$ values escalate. 

\begin{figure}[!htp]
\begin{center}
\includegraphics[width=1\linewidth]{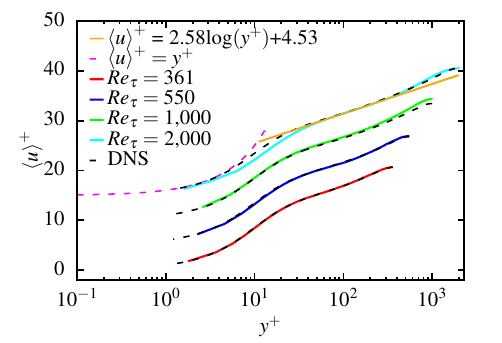}
\caption{Comparison of $\left<u\right>^+$ as function of $y^+$ for different values of $Re_\tau$ {and the WALE subgrid model.} $u_{\text{rms}}^+$ profiles are plotted with an offset of $\Delta \left < . \right >^+ = 5$ for clarity. DNS data is shown with dashed black lines. The dashed pink and solid orange lines depict the universal behavior of the turbulent streamwise velocity profile.}
\label{fig: 15 Smooth mean flow profiles for different Reynolds}
\end{center}
\end{figure}
In Fig. \ref{fig: 15 Smooth mean flow profiles for different Reynolds} (a), we present mean velocity profiles computed with LES-WALE for $Re_\tau=361$, $550$, $1,000$, and $2,000$ using new meshes M4, G21, G34, and G41.
A good agreement is observed between LES-WALE and DNS, particularly in the inner layer and overlap regions, with the data gradually converging to a logarithmic distribution. This log-law is visually fitted as $\left< u\right>^+ = (1/{\kappa})\text{log}(y^+)+B$, with coefficients ${\kappa}=0.387$ and $B=4.53$ from Pirozzoli \textit{et al.} \citep{pirozzoli2021one}. The results closely aligns with the estimates from direct fitting of mean velocity profiles in pipe flows \citep{marusic2013logarithmic}, yielding $\left< u\right>^+ = (1/{\kappa})\text{log}(y^+)+B$, with ${\kappa}=0.391$ and $B=4.34$. Not much difference is obtained compared to results from coarser meshes, shown in Fig.~\ref{fig: 15 Bad Smooth mean flow profiles for different Reynolds} (a). 
\begin{figure*}[!htp]
\begin{center}
\subfloat[]{
\includegraphics[width=0.45\linewidth]{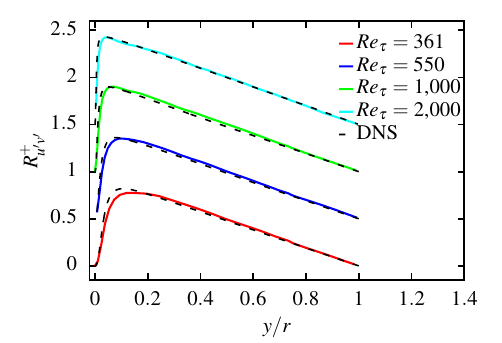}}
\subfloat[]{
\includegraphics[width=0.45\linewidth]{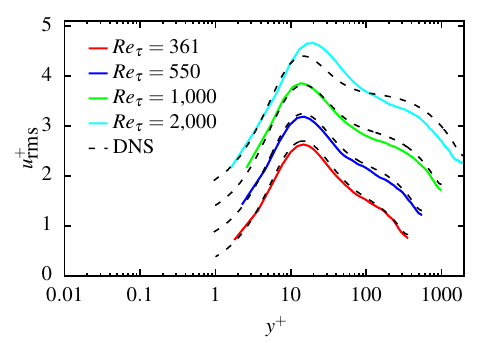}}

\subfloat[]{
\includegraphics[width=0.45\linewidth]{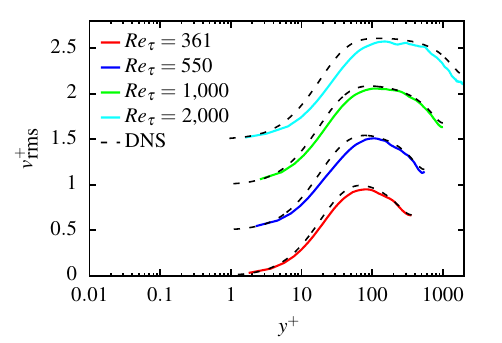}}
\subfloat[]{
\includegraphics[width=0.45\linewidth]{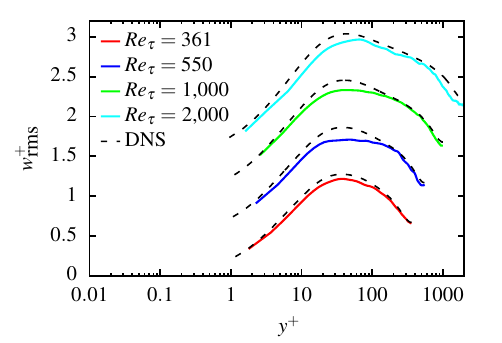}}
\caption{(a) Turbulent shear stresses, $R^+_{u'v'}$ and turbulent velocity fluctuations (b) $u_{\text{rms}}^+$, (c) $v_{\text{rms}}^+$, and (d) $w_{\text{rms}}^+$, corresponding to different values of $Re_\tau$ {and the WALE subgrid model}, are shown as a function of $y^+$ and offset by $\Delta \left < . \right >^+ = 1$ for clarity. In all subfigures, the DNS data are represented in dashed black lines.}
\label{fig: 15aa Smooth mean flow profiles for different Reynolds}
\end{center}
\end{figure*}

Figure \ref{fig: 15aa Smooth mean flow profiles for different Reynolds} (a) provides insights into the turbulent shear stress distribution for varying $Re_\tau$ values. As $Re_\tau$ increases, the shear stress profiles flatten, with the peak value rising toward unity, and its position shifting further from the wall. This well-established behavior \citep{lee2015direct} is characterized by the asymptotic relationship that accounts for the position of the turbulent shear stress peak, $y_p^+$ \citep{afzal1982fully}, 
\begin{equation}
y^+_p \simeq \sqrt{\frac{Re_\tau}{{\kappa}}}.
\label{eq: RSS peak location}
\end{equation} 

The precision of Eq. \eqref{eq: RSS peak location}, with ${\kappa}=0.387$, is notably accurate starting from $Re_\tau \simeq 1,000$ onwards, indicating the onset of a near-logarithmic layer \citep{chin2014reynolds}. In our simulations, the predicted peak position, $y^+_p$, closely aligns with expected values for $Re_\tau=1,000$ and $2,000$, with deviations of approximately $5\%$. However, for $Re_\tau < 1,000$, the predictions deviate significantly from the asymptotic relation, showing deviations of roughly $27\%$. 

Based on the proposed criteria we have further looked upon the velocity fluctuations results for $Re_\tau=361, 550, 1000$ and $ 2,000$ in Fig. \ref{fig: 15aa Smooth mean flow profiles for different Reynolds} (b-d). Compared with the results obtained using Taylor scaling (Fig.~\ref{fig: 15 Bad Smooth mean flow profiles for different Reynolds} (b)), the present results matches with the DNS, as shown in Fig.~\ref{fig: 15aa Smooth mean flow profiles for different Reynolds}. Slight discrepancies between DNS results and LES at $Re_\tau=2,000$ can still be observed; nevertheless, both the primary peaks and the secondary overshoot region are well captured in comparison to existing literature~\citep{berrouk2007stochastic,ferrer2019high}. Unlike previous results, the mesh requirement for $Re_\tau = 2000$ is $\sim 220$M cells and increases {tremendously} with $Re_\tau$. Therefore, we restrict this study to $Re_\tau=2000$. 

\subsection*{Single point statistics}
To quantitatively depict near-wall regions at different $Re_\tau$, we utilize quadrant analysis for streamwise and wall-normal velocity fluctuations. Quadrant analysis divides the ($u', v'$) plane into four quadrants ($Q_1, Q_2, Q_3, Q_4$) based on sign combinations of velocity fluctuations. Specifically, $Q_1$ corresponds to positive values of both $u'$ and $v'$, $Q_2$ represents $u'<0, v'>0$, $Q_3$ indicates $u'<0, v'<0$, and $Q_4$ corresponds to $u'>0, v'<0$. Ejection-like events occur in $Q_2$, while sweep-like events are in $Q_4$. Quadrant analysis provides insights into the dynamics of near-wall streaks, initially observed by Kline \textit{et al.} \citep{kline1967structure} and further explored by Corino \& Brodkey \citep{corino1969visual} and Wallace \textit{et al.} \citep{wallace1972wall}. Willmarth \& Lu \citep{willmarth1972structure} computed quadrant contributions to the total shear stress, highlighting the importance of ejection and sweep events in the production of turbulent Reynolds shear stress. The shear stress plays a crucial role in turbulence dynamics, appearing in the LES equation and the turbulent kinetic energy equation's production term, $P_{ij}$, multiplied by the mean shear stress.
\begin{equation}
P^k_{ij} = -\left<u'_iu'_j\right>\dfrac{\partial\left<u_i\right>}{\partial x_j}.
\label{eq:production}
\end{equation}
\begin{figure*}[!htp]
\begin{center}
\subfloat[]{
\includegraphics[width=0.35\linewidth]{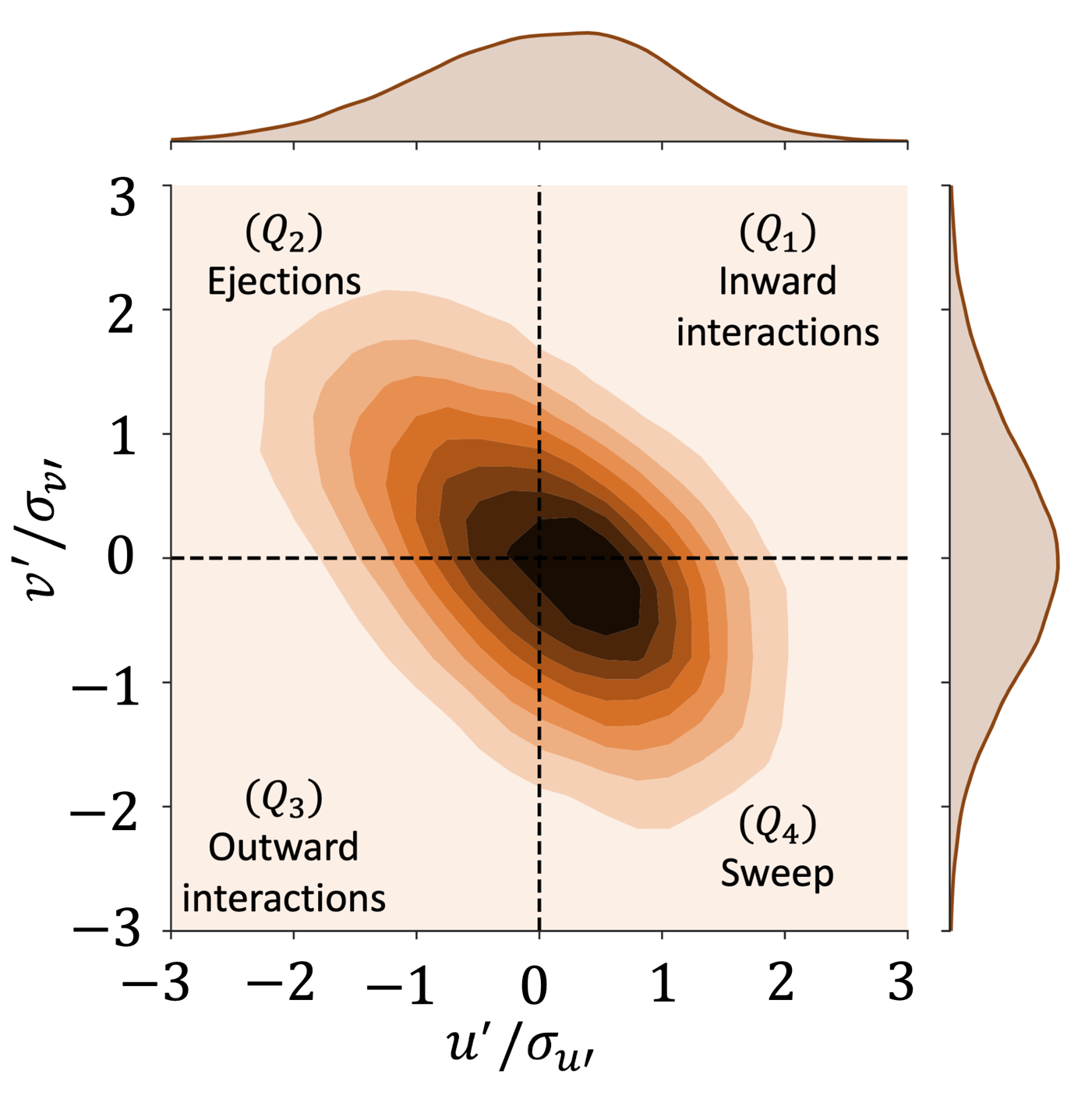}}
\subfloat[]{
\includegraphics[width=0.35\linewidth]{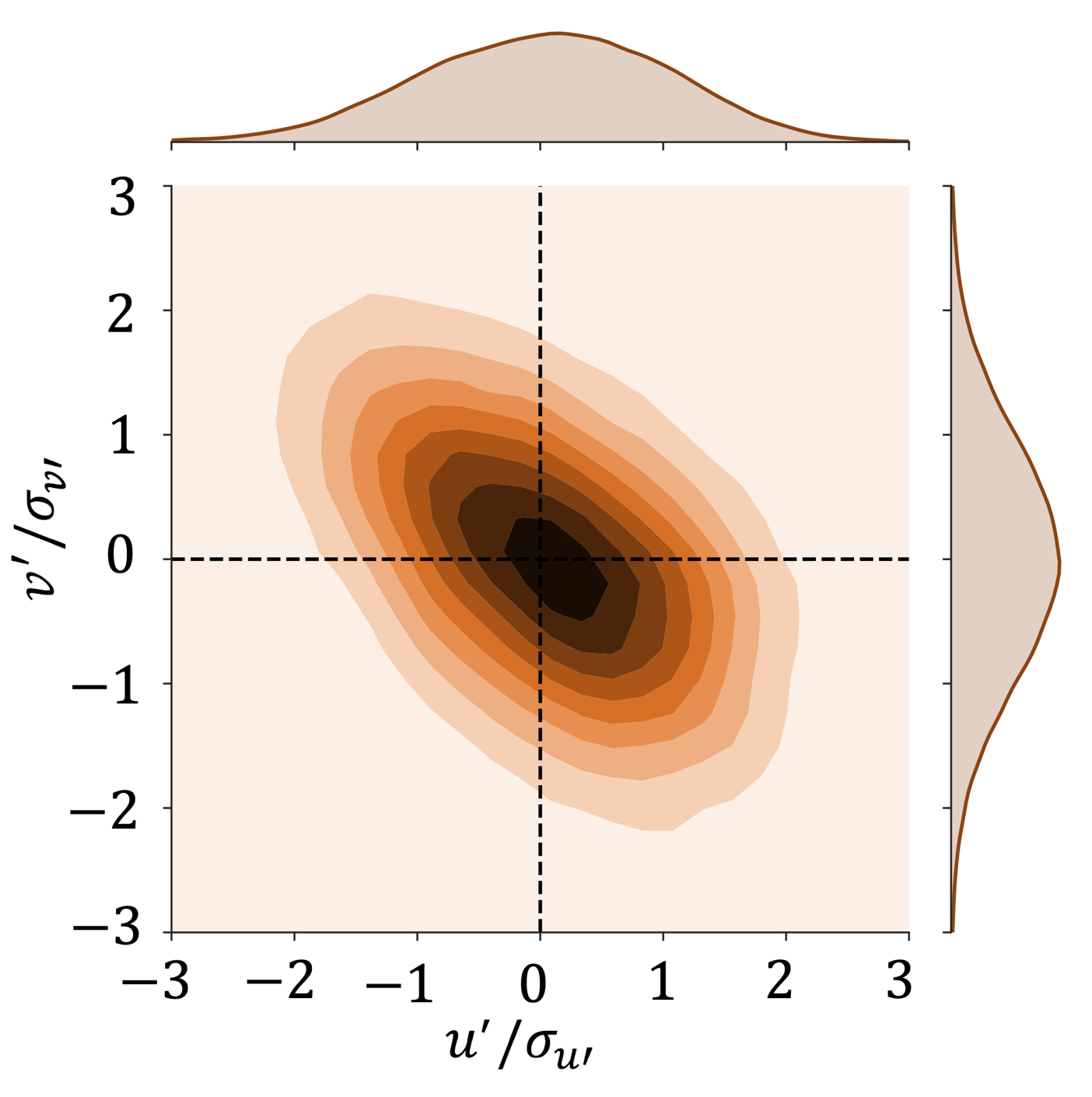}}

\subfloat[]{
\includegraphics[width=0.35\linewidth]{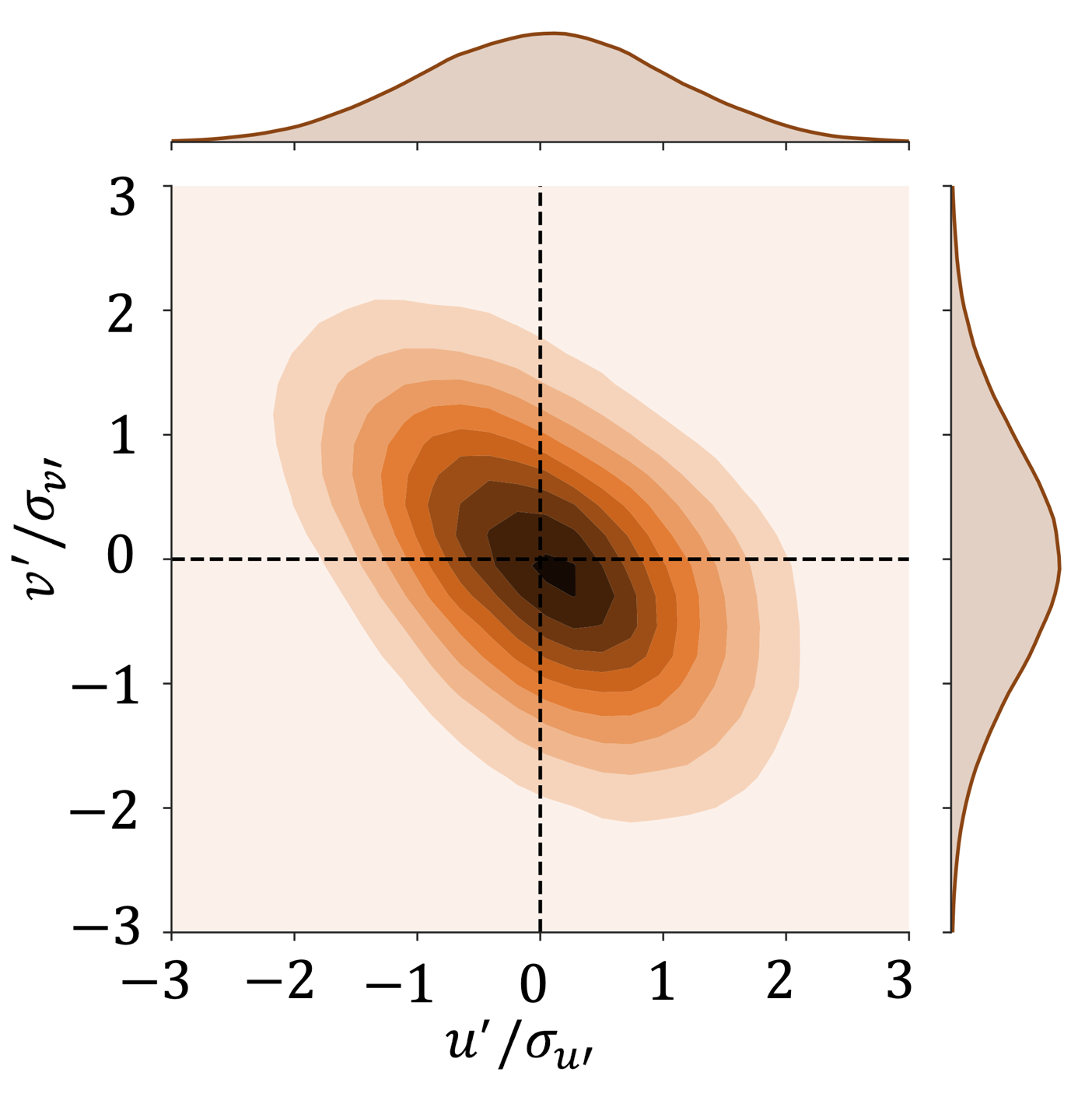}}
\subfloat[]{
\includegraphics[width=0.35\linewidth]{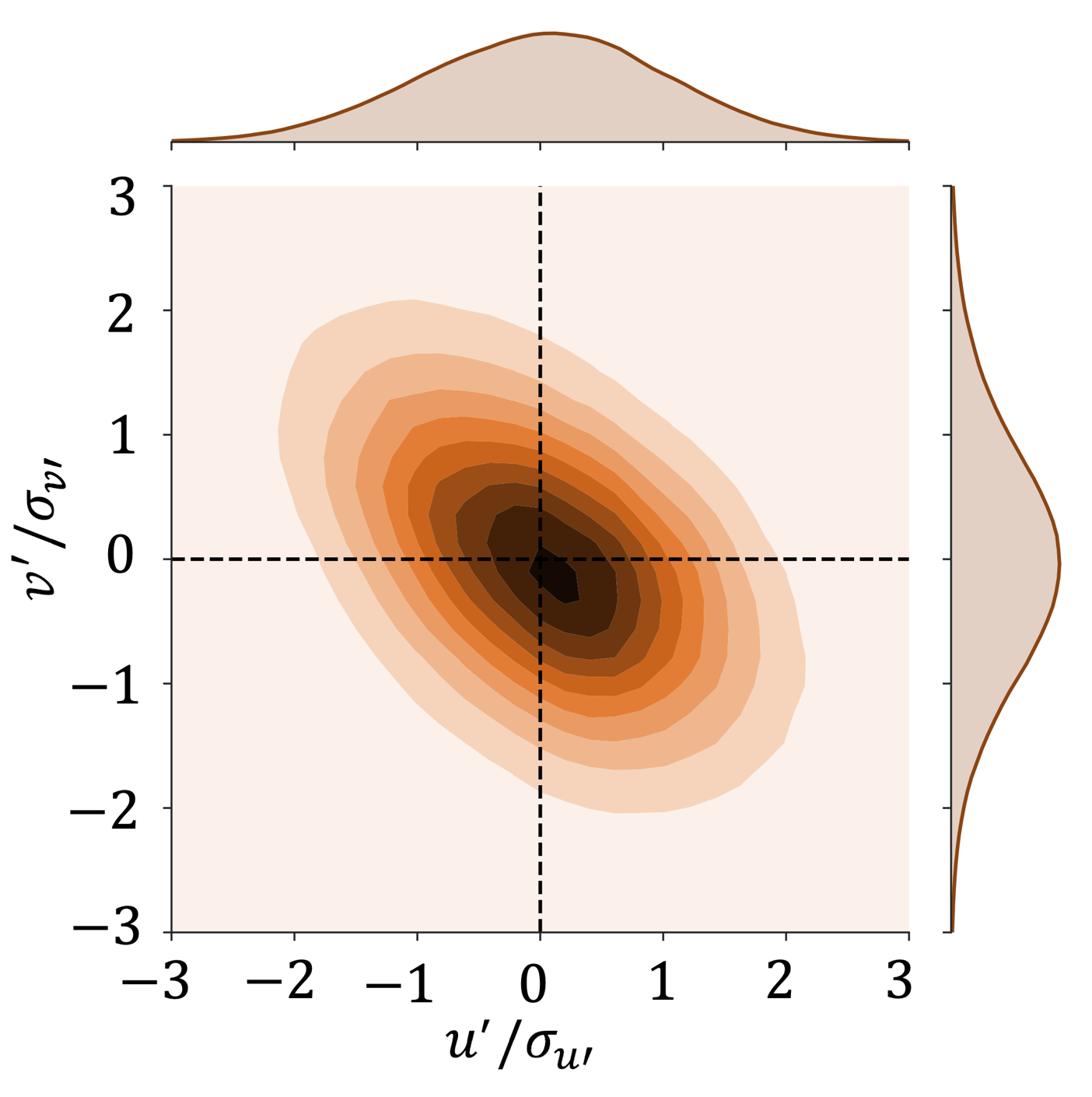}}
\caption{Quadrant plots illustrate axial, $u'$, and wall-normal, $v'$, velocity fluctuations normalized by their root-mean-squares, $\sigma_{i'}$, for different turbulent $Re_\tau$ and for the WALE subgrid model, at $y^+ = 3\sqrt{Re_\tau}$: (a) $Re_\tau=361$, (b) $Re_\tau=550$, (c) $Re_\tau=1,000$, and (d) $Re_\tau=2,000$. Each plot includes PDFs of $u'/\sigma_{u'}$ and $v'/\sigma_{v'}$ on the top and right corners, respectively.}
\label{fig: 16b JPDFS for smooth pipe flows for different Re using contours}
\end{center}
\end{figure*}
\begin{figure*}[!htp]
\begin{center}
\subfloat[$y^+ = 5$]{
\includegraphics[width=0.4\linewidth]{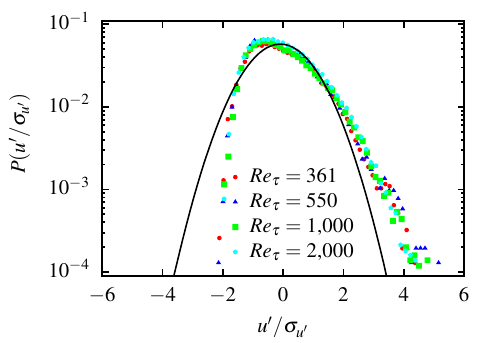}}
\subfloat[$y^+ = 150$]{
\includegraphics[width=0.4\linewidth]{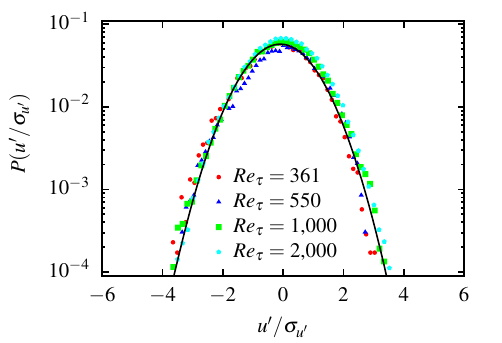}}

\subfloat[$y^+ = 5$]{
\includegraphics[width=0.4\linewidth]{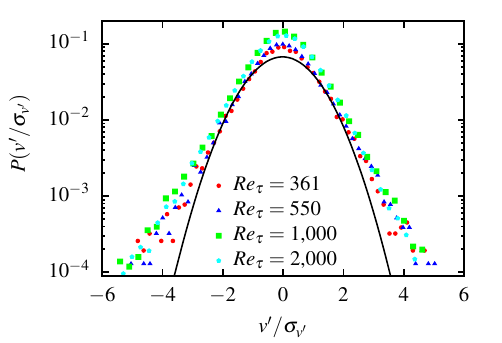}}
\subfloat[$y^+ = 150$]{
\includegraphics[width=0.4\linewidth]{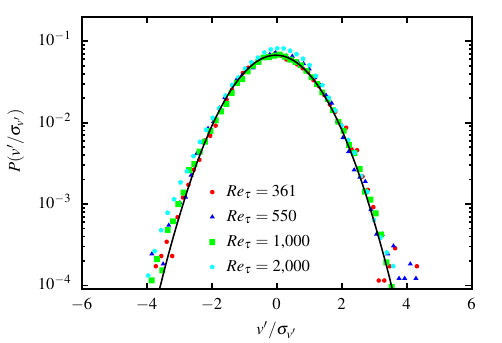}}
\caption{Probability Distribution Functions (PDF) of streamwise velocity fluctuations,  $u'/\sigma_{u'}$ (a) close to the wall ($y^+ = 5$), (b) in the outer region ($y^+ = 150$), and wall-normal velocity fluctuations, $v'/\sigma_{v'}$ (c) close to the wall ($y^+ = 5$), (d) in the outer region ($y^+ = 150$), {for different values of $Re_\tau$ and the WALE subgrid model.} The PDF are normalized by their rms values, $\sigma_{i'}$.}
\label{fig: 18 PDFS and spectras for smooth pipe flows for different Re}
\end{center}
\end{figure*}

Fig. \ref{fig: 16b JPDFS for smooth pipe flows for different Re using contours} shows contour plots of the joint probability density function (JPDF) of velocity fluctuations, $P(u',v')$, normalized by their local rms values, $\sigma_{u'}$ and $\sigma_{v'}$. These plots are at $y^+=3\sqrt{Re_\tau}$ for four $Re_\tau$ values. The JPDF iso-contours have an elliptical shape, mainly spanning quadrants Q$_2$ and Q$_4$, with the maximum value in Q$_4$, indicating a higher frequency of $\left\{u'v'(t) < 0\right\}$ events, particularly with more extreme values in Q$_2$ (ejections). Normalized PDFs are also displayed on the top ($u'$) and right ($v'$) corners in each plot. Fig. \ref{fig: 18 PDFS and spectras for smooth pipe flows for different Re} presents PDFs of streamwise and wall-normal velocity fluctuations, $P(u')$ and $P(v')$, at various distances from the wall for different $Re_\tau$. These PDFs concern fluctuations relative to the mean and are normalized by the root-mean-square value. In the near-wall region (Fig. \ref{fig: 18 PDFS and spectras for smooth pipe flows for different Re} (a)), the distribution of streamwise velocity fluctuations, $u'$, deviates significantly from a Gaussian shape, indicating intermittency with more intense and intermittent high-speed streaks close to the wall. This behavior contributes to the positive skewness observed in experiments \citep{ciclope2017} and DNS \citep{redjem2007direct}. The PDFs of wall-normal fluctuations, $v'$, (Fig. \ref{fig: 18 PDFS and spectras for smooth pipe flows for different Re} (c)), are weakly asymmetric, favoring descents over ascents. In the outer region of the pipe ($y^+ = 150$), the PDF shapes approach a Gaussian distribution (Fig. \ref{fig: 18 PDFS and spectras for smooth pipe flows for different Re} (b), (d)), indicating that $Re_\tau$ has minimal impact on the overall dominance of ejection- and sweep-like events.
\subsection*{Anisotropy of turbulence for different Reynolds numbers}
\begin{figure*}[!htb]
\begin{center}
\subfloat[$Re_\tau=361$]{
\includegraphics[width=0.35\linewidth]{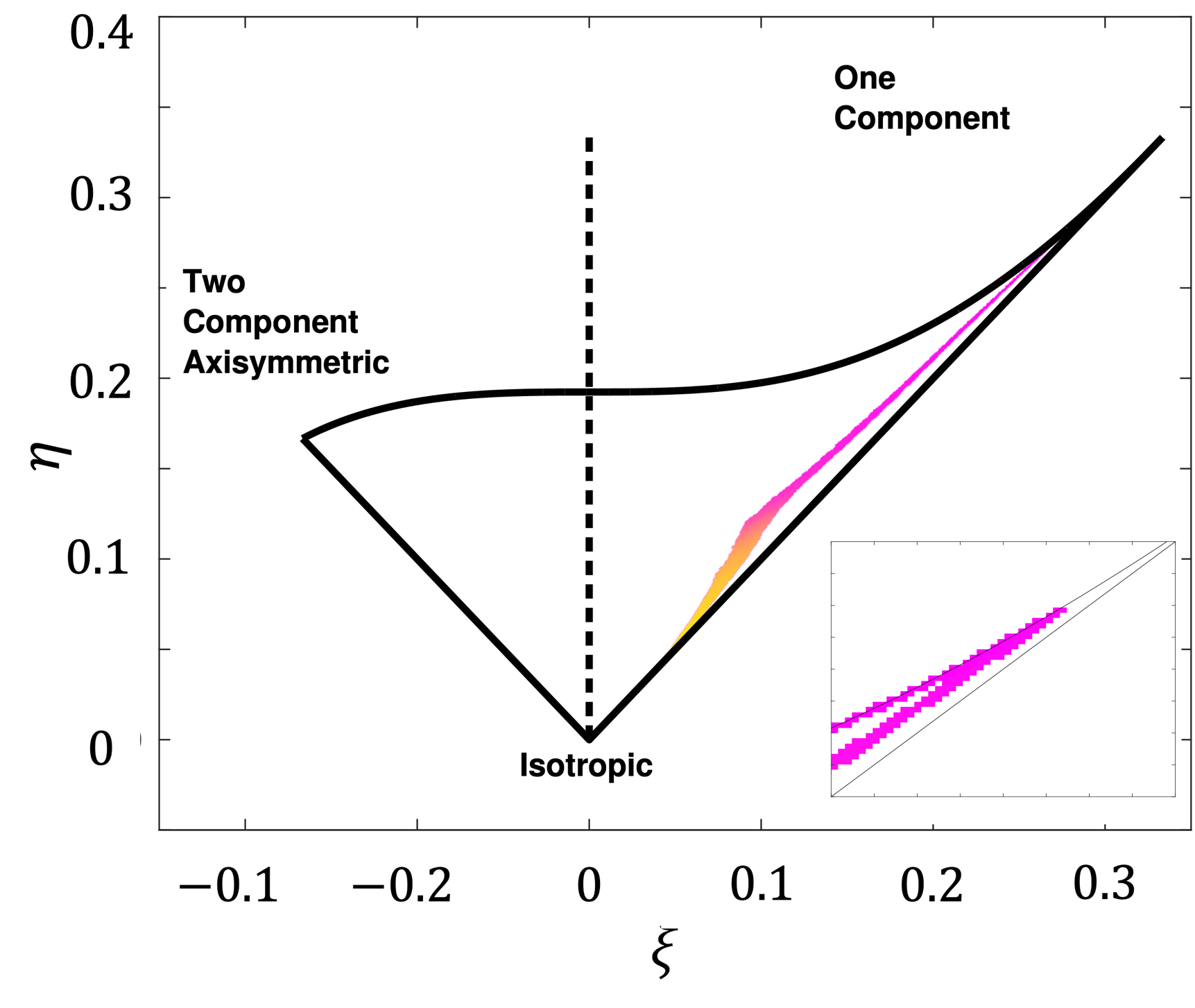}}
\subfloat[$Re_\tau=550$]{
\includegraphics[width=0.35\linewidth]{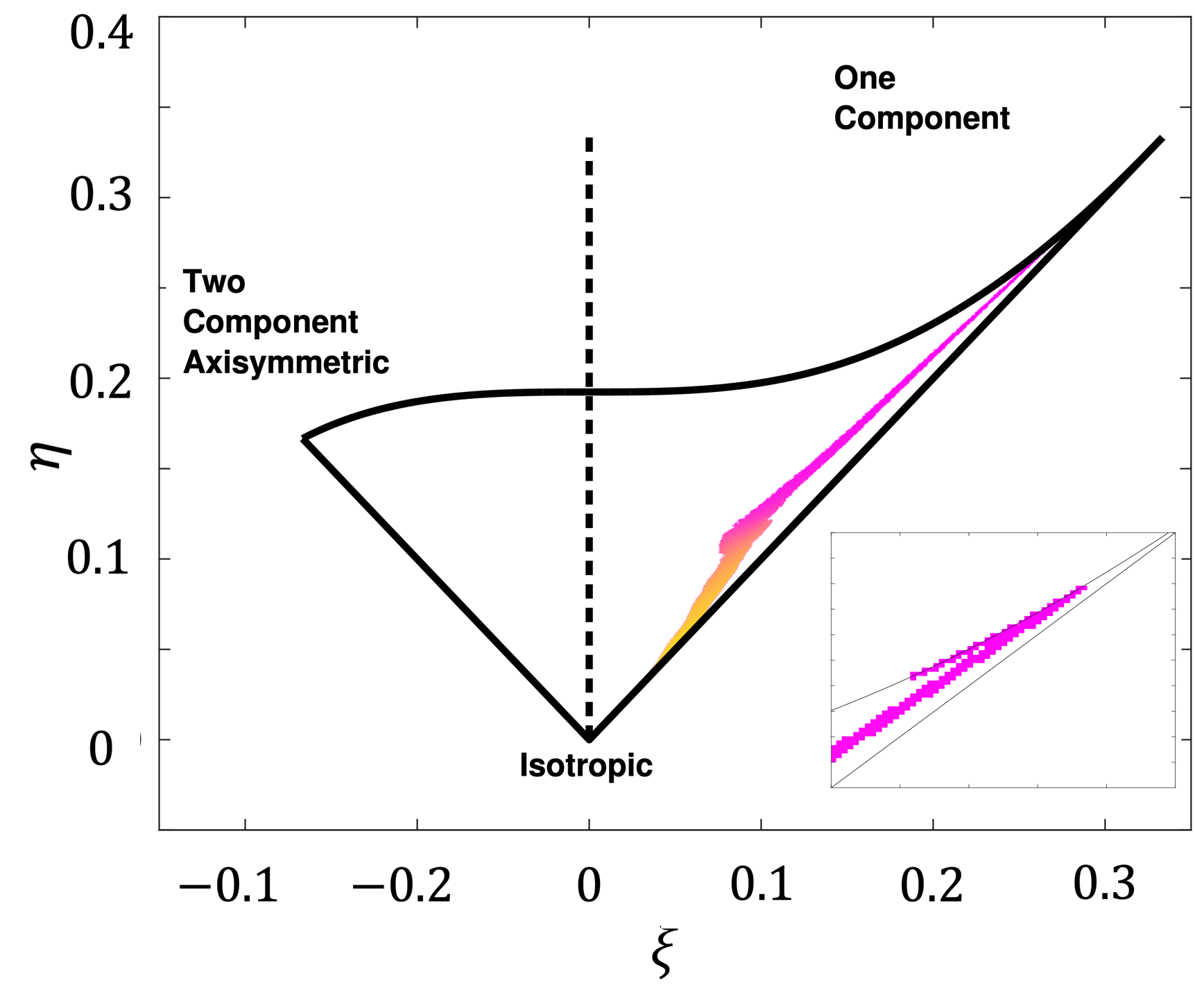}}

\subfloat[$Re_\tau=1,000$]{
\includegraphics[width=0.35\linewidth]{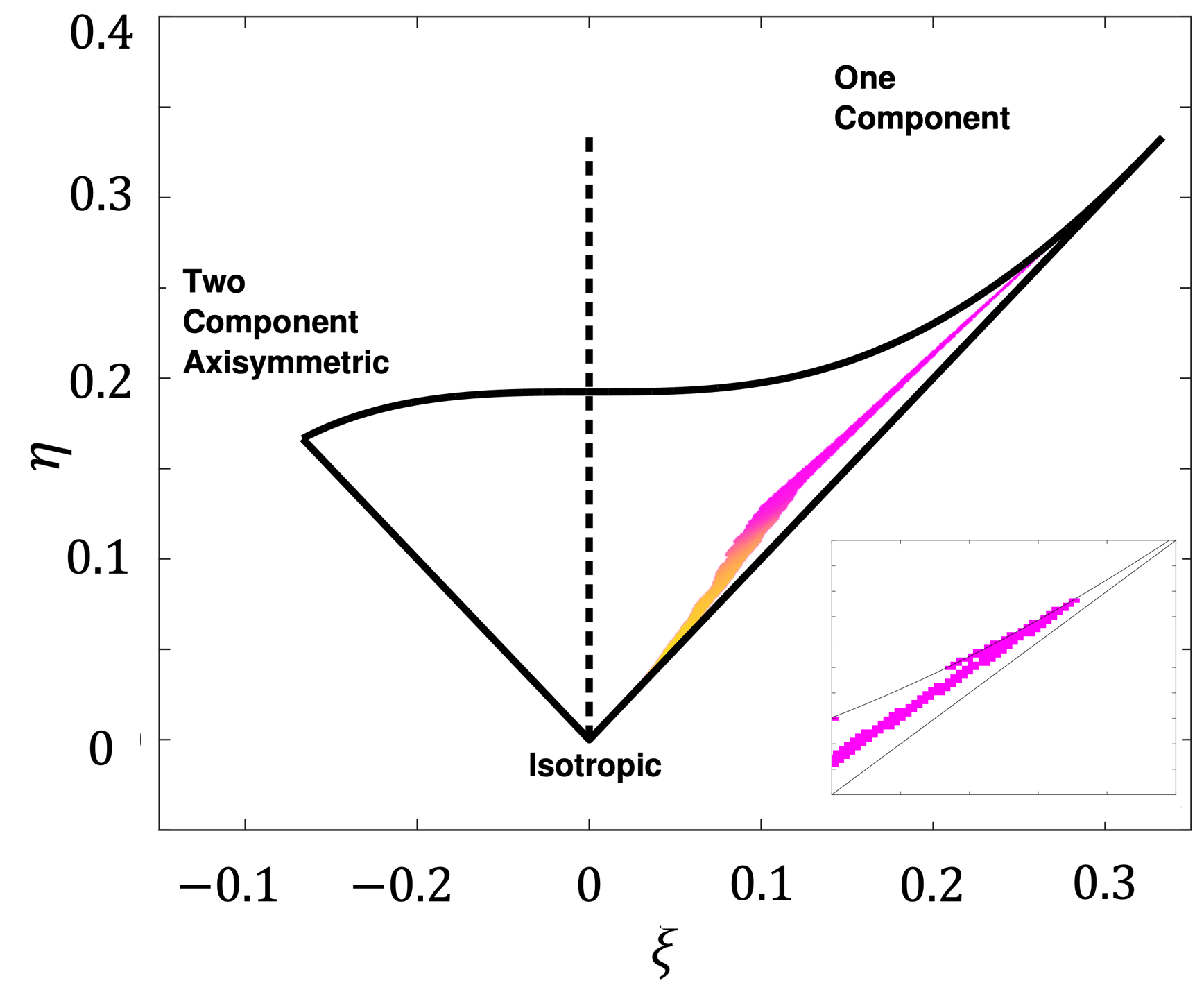}}
\subfloat[$Re_\tau=2,000$]{
\includegraphics[width=0.35\linewidth]{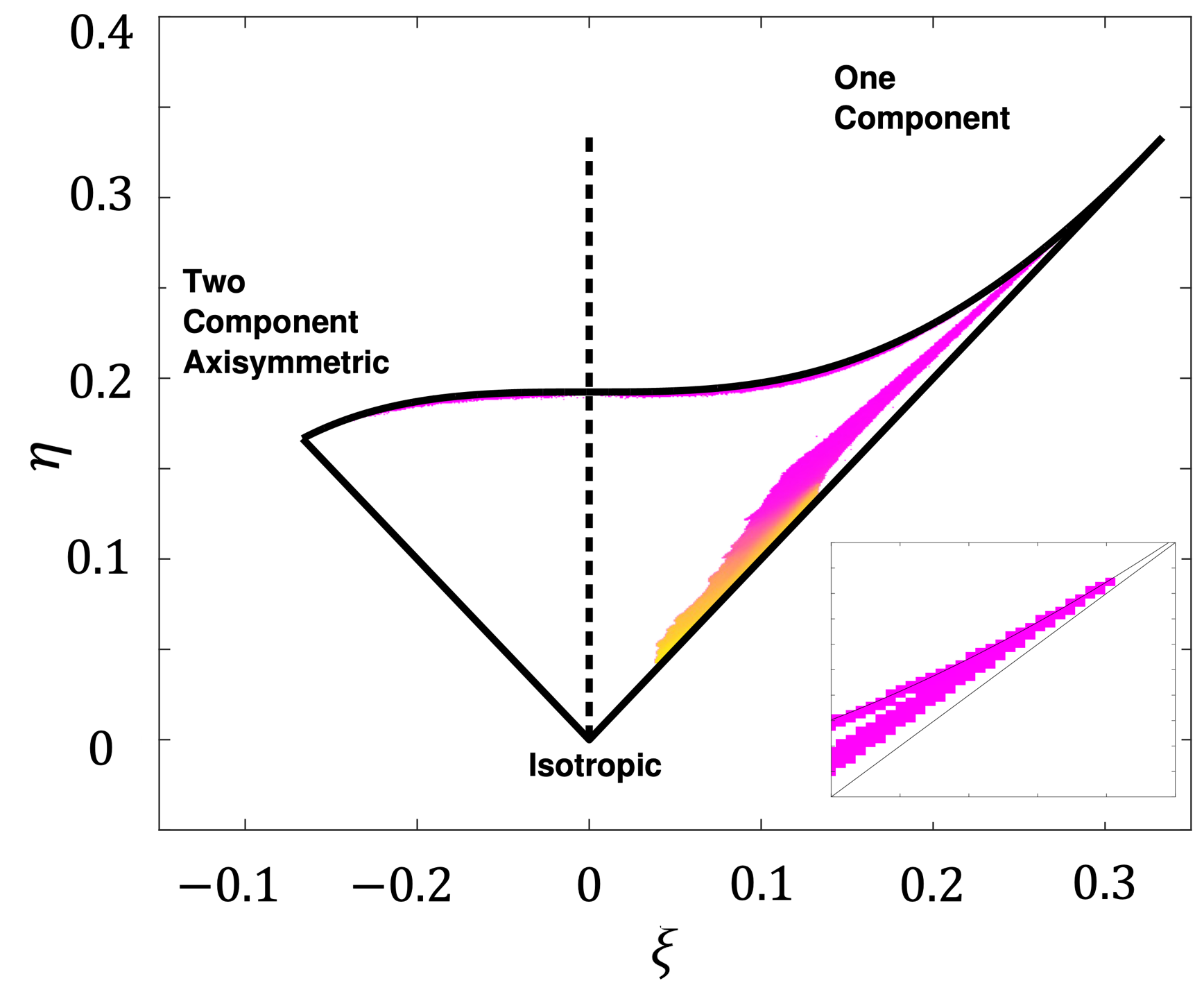}}
\caption{Anisotropy-invariant mapping of turbulence from present LES-WALE data at (a) $Re_\tau=361$, (b) $Re_\tau=550$, (c) $Re_\tau=1,000$, and (d) $Re_\tau=2,000$. Data points for each subgrid model are based on all cells in the domain and are colored with normalized wall distance values, $y^+$. The color map varies from pink to yellow. Insets in all subfigures show the same zoom of the one-component turbulence area.}
\label{fig: 20 Anisotropy maps for different Re}
\end{center}
\end{figure*}

{The anisotropy of unsteady turbulent flows is important characteristics of turbulence. Turbulence anisotropy is represented by the degree of departure from isotropic turbulence. Accurate numerical simulations require precise predictions of anisotropy for various flow situations such as those encountered in atmospheric science and engineering design}, which can be quantified using the second ($II_b$) and third ($III_b$) invariants of the normalized anisotropy tensor, $b_{ij}$, given by
\begin{equation}
b_{ij} = \frac{\left< \overline{ u'_iu'_j}\right>}{\left< \overline{ u'_ku'_k}\right>} - \frac{1}{3} \delta_{ij}.
\label{anisotropy}
\end{equation}
The state of anisotropy can then be characterized with the two variables $\eta$ and $\xi$ defined as
\begin{equation}
\eta^2 = -\frac{1}{3}II_b
\end{equation}
and
\begin{equation}
\xi^3 = -\frac{1}{2}III_b.
\end{equation}
Reynolds stress tensor states are confined within the Lumley triangle, defined in the $\left(\xi,\eta\right)$ plane, with different realizable states characterized by the invariants of the normalized anisotropy tensor, $b_{ij}$. Details on these states can be found in \cite{lumley1977return}. Anisotropy maps in Fig. \ref{fig: 20 Anisotropy maps for different Re} depict variations for four $Re_\tau$ values across $y^+$. These maps utilize all somain cells, and post-processed values of $\eta$ and $\zeta$ are represented with a color map of $y^+$, indicating the likelihood of specific turbulent states. Near the wall, turbulence is highly two-dimensional (dominated by $u'u'$ and $w'w'$ fluctuations), transitioning toward a one-dimensional, one-component state as $u'u'$ peaks around $y^+ \simeq 8$, as highlighted in the inset of Fig. \ref{fig: 20 Anisotropy maps for different Re}. As $y^+$ increases, data points move toward the origin, aligning along the axisymmetric state, signifying an approach to isotropy. These findings align well with Pope's analysis \cite{pope2000turbulent} of DNS data from Kim \textit{et al.} \cite{kim1987turbulence}, where two-component turbulence was identified near the wall ($y^+ < 5$), transitioning to axisymmetric expansion in the channel's remaining portion. The anisotropy maps exhibit a universal behavior across different $Re_\tau$ values.
%
\section{Conclusions}
The present work assesses the prediction capabilities of wall-modeled  Large Eddy Simulations (LES) at moderate-to-high Reynolds numbers and in the limit of coarse meshes, in the OpenFOAM framework. Initially, results for a turbulent pipe flow at $Re_\tau=361$ were compared against DNS data available in the literature. The simulation results obtained using the WALE subgrid model showed excellent qualitative and quantitative prediction accuracy when compared to DNS data; nonetheless, with slight underprediction of all turbulence characteristics. We further assessed the accuracy of the WALE subgrid model by performing a grid resolution study on four different grids (M1, M2, M3, and M4, where M1 is the coarsest grid, and  M4 is the finest grid). The results showed a gradual increase in accuracy of zeroth-, first-, and second-order critical turbulence statistics with grid resolution. However, as a trade-off between numerical accuracy and computational wall clock time, we decided to proceed with a rather coarse M2 grid for further detailed investigations. 

Based on the grid resolution study, we performed LES for five different subgrid models: the Wall-Adapting Local Eddy viscosity model (WALE), the Smagorinsky model (SMG), the One-Equation Eddy Viscosity Model (OEEVM), the Localized Dynamic Kinetic energy Model (LDKM), and the Differential Stress Equation Model (DSEM) on grid M2. The results obtained demonstrated that WALE, LDKM, and DSEM were able to reproduce the main and most critical characteristics of turbulent flows, featuring regions of high strain rates and large-scale vortices, with higher accuracy than OEEVM and SMG, both quantitatively and qualitatively, in comparison to DNS data from the literature. Observing the flow statistics, it could be evidenced that the estimates of the mean flow profiles and turbulent Reynolds shear stresses were consistent for the three subgrid models (WALE, LDKM, and DSEM). Still, DSEM showed a better overall performance in predicting the near-wall and outer region statistics, particularly for $v^+_{\text{rms}}$ and $w^+_{\text{rms}}$, possibly due to its anisotropic assumption. The DSEM model still slightly overpredicts all the flow statistical quantities in the core region. This is possibly due to the fact that the physical constants of this model are not computed dynamically, as opposed to the LDKM model whose accuracy is then improved. However, the differences between the results obtained with WALE and LDKM were insignificant, and both models provided highly accurate profiles of mean flow and turbulent Reynolds shear stresses. 

Additionally, to strengthen our analysis and assess the prediction capabilities of subgrid models, we performed LES utilizing WALE for four moderate turbulent $Re_\tau=361$, $550$, $1,000$, and $2,000$, and compared them against DNS data available in the literature.
To stress on the potential of the near-wall-modeled LES in the limit of a coarse grid and high $Re_\tau$, in the present analysis the M2 grid is chosen as a base grid for $Re_\tau = 361$. The grid has then been refined to resolve roughly the same amount of turbulent shear stresses and turbulent kinetic energy  for $Re_\tau = 550$ to $2,000$.
Estimating the zeroth- and first-order statistics showed that WALE gives reasonable results compared to DNS upto $Re_\tau=550$ and deviate slowly with a further increase of $Re_\tau$. In the present wall-modeled LES another outcome that surprised us is the failure of the wall models for high $Re_\tau$.   
Firstly, the Reynolds stress tensors exhibited reasonable agreement with DNS data in the outer region for all values of $Re_\tau$. However, they also revealed a dramatic deviation from the DNS data in the sub-viscous layer region at all values of $Re_\tau$.
Secondly, velocity fluctuation profiles show strong deviations from DNS, even for moderate $Re_\tau$. Surprisingly, the secondary-peak region is poorly captured. This observation is peculiar and raises questions about the validity of the scaling based on Taylor and integral scales ($\lambda/l_I \propto Re^{-1/2}$) used to obtain relatively similar meshes for different values of $Re_\tau$. Based on the considered scaling, similar resolution was expected for different $Re_\tau$ values, but the results showed the opposite both in the wall region (a smaller number of mesh nodes when moving from $Re_\tau = 1,000$ to $2,000$) and outer region (secondary peak is not captured). The results from Berrouk \textit{et al.} and Ferrer \textit{et al.} also demonstrate discrepancies with DNS data, both regarding the primary peak and the secondary overshoot.
This observation suggests that LES with WALE requires greater refinement in the wall-normal direction than the one used in the current work to more accurately anticipate the behavior of the smallest-scale motions. This indi
cates the need for a different scaling factor to resolve a relatively similar amount of turbulent shear stress with increasing $Re$. Upon further investigation of grid convergence using Kolmogorov scaling, it was found that to accurately capture second-order statistics, particularly secondary overshoots in fluctuations, it is necessary to utilize $\Delta x^+ \leq 20-25$ and $\Delta z^+ \leq 10$. This leads to an actual number of grid points required for well-resolved LES of pipe flow, denoted as $N_{LES}$, at approximately $(2.25 \cdot Re_\tau)^{9/4}$. The coefficient differs from DNS, which is $3$. While the computational savings compared to DNS are extremely significant, they amount to at least 30\%.

To conclude, our findings suggest that the scaling strategy we have explored in this paper is not effective, and we advise against others attempting to replicate this approach. Rather, we recommend a different methodology that involves a direct measurement of the resolution using appropriate criteria such as modeled kinetic energy and grid variations that ensure comparable resolutions across different $Re_\tau$. Although this approach can be time-consuming and challenging, it is essential for achieving relatively similar resolutions across different Reynolds numbers. Therefore, we suggest either the use of grid spacing constraint proposed above or the use of simulation methods that replace the filter width with a modeled length scale and include the modeled kinetic energy for resolution calculations. 
By adopting these methods, one can ensure a more accurate and reliable resolution for the simulations in various physical and engineering applications.
\section*{Acknowledgement}
The authors greatly appreciate the financial support provided by Vinnova under project number "2020-04529". Computer time was provided by the Swedish National Infrastructure for Computing (SNIC), partially funded by the Swedish Research Council through grant agreement no. 2018-05973. 
\section*{Author declarations}
\subsection*{Conflict of interest}
The authors have no conflicts of interest to disclose.
\subsection*{Data availability}
The data that support the findings of this study are available from the corresponding author upon reasonable request.
\appendix 
\section{Mean velocity-defect profile}\label{Appendix 0}
\begin{figure}[H]
\begin{center}
\includegraphics[width=0.8\linewidth]{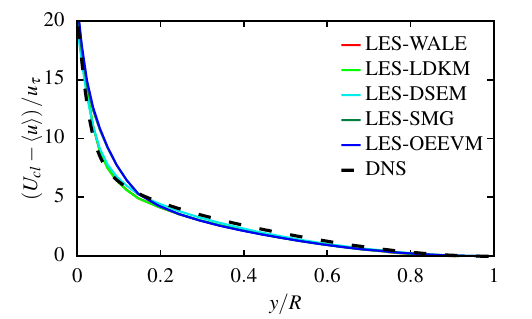}
\caption{{Mean velocity-defect profiles for $Re_\tau=361$ and five different subgrid models, normalized by $u_\tau$. The value of $u_\tau$ is estimated from the simulations. $U_{cl}^+$ is the normalized centerline velocity. The dashed black line presents DNS data from El Khoury \textit{et al.} \citep{el2013direct}.}}
\label{fig: 4b_rebuttal}
\end{center}
\end{figure}

{Figure~\ref{fig: 4b_rebuttal} shows mean velocity-defect velocity profiles in the outer region for various subgrid models. All resuts are normalized with $u_\tau$. The difference between Fig.~\ref{fig: 6 Re361 Mean Flow Profiles subgrid models}(a) and Fig.~\ref{fig: 4b_rebuttal} lies in the normalization. In Fig.~\ref{fig: 6 Re361 Mean Flow Profiles subgrid models}(a), all results are normalized by $u_\tau^0$. For the definition, refer to Table.~\ref{table: friction factor subgrid dependence}.} 
\section{Autocorrelations functions}\label{Appendix I}
The correlations of the velocity fluctuations in each direction are typically characterized by means of statistical tools. Notably, a standard tool to assess the relationship between the velocity fluctuations in the streamwise and wall-normal directions is the coefficient of correlation. Our measurements allow us to quantitatively assess the correlations of velocity fluctuations in time and space. The temporal correlation functions of the velocity fluctuation components, defined as
\begin{equation}
C_{i'}(T) = \frac{\left< i'(t+T)i'(t) \right>_t}{\left< \left[ i'(t) \right]^2 \right>_t},\quad i= u,v,
\label{eq:16}
\end{equation}
where the brackets indicate the time average and $T$ is the time interval between two consecutive sample points. 

Similarly, the spatial cross-correlation function of each velocity component is defined as
\begin{equation}
C_{i'}(X) = \frac{\left< i'(x+X)i'(x) \right>_x}{\left< \left[ i'(x) \right]^2 \right>_x},\quad i= u,v,
\label{eq:17}
\end{equation}
where the brackets indicate an average over the $x$-direction and $X$ is the distance between two points along the given direction.  

\begin{figure}[H]
\begin{center}
\subfloat[]{
\includegraphics[width=0.5\linewidth]{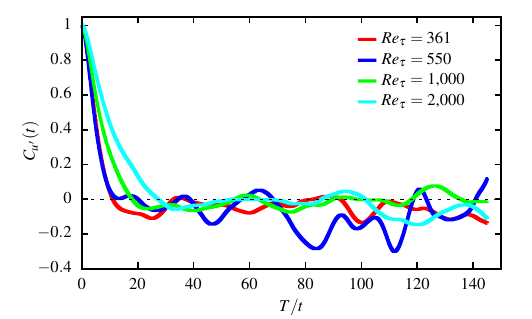}}
\subfloat[]{
\includegraphics[width=0.5\linewidth]{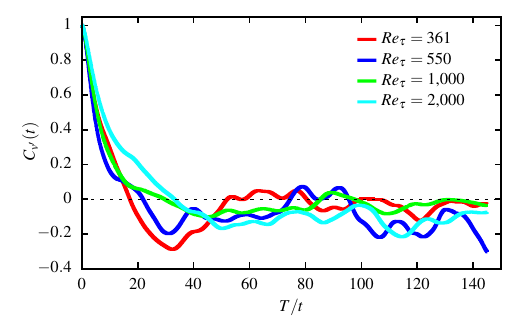}}

\subfloat[]{
\includegraphics[width=0.5\linewidth]{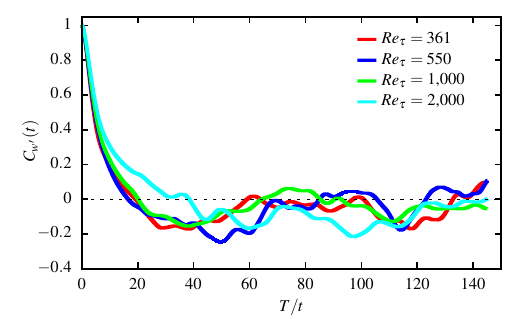}}
\subfloat[]{
\includegraphics[width=0.5\linewidth]{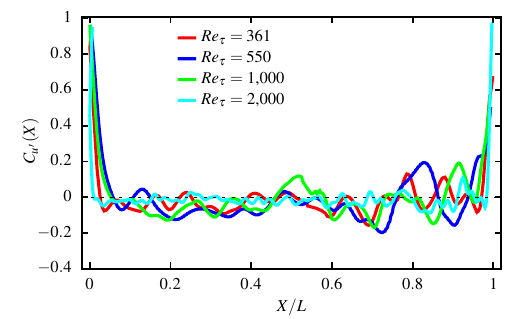}}

\subfloat[]{
\includegraphics[width=0.5\linewidth]{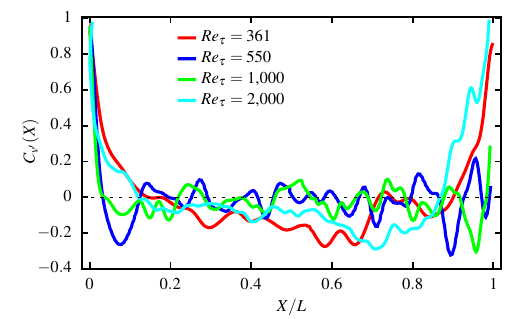}}
\subfloat[]{
\includegraphics[width=0.5\linewidth]{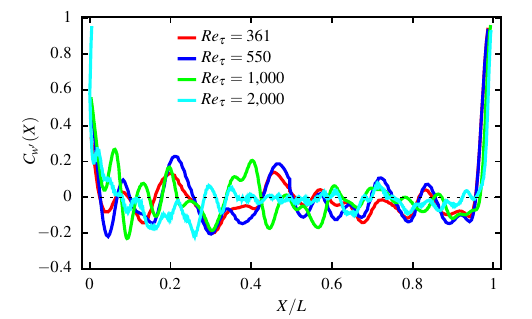}}
\caption{ Normalized one-dimensional temporal and spatial autocorrelation functions of (a), (d) axial velocity fluctuations, $u'$, (b), (e) wall-normal velocity fluctuations, $v'$, and (c), (f) azimuthal velocity fluctuations, $w'$, for different values of $Re_\tau$.}
\label{fig: 19 autocorrelations for smooth pipe flows}
\end{center}
\end{figure}

The correlations of the two-point velocity fluctuations computed from Eq. \eqref{eq:16} in the temporal and spatial domains as functions of the turbulent $Re_\tau$ are shown in Fig. \ref{fig: 19 autocorrelations for smooth pipe flows} (a-c) and \ref{fig: 19 autocorrelations for smooth pipe flows} (d-f), respectively. It should be noted that data for cross-correlation analysis in the $x$-direction are extracted at $y=R$ and averaged using 100 line probes. The profiles indicate that the chosen duration of simulation is long enough to capture all the relevant turbulent structures, since they fluctuate around the zero value for all components. The corresponding correlation times, computed as $t_{i'} = \int_0^{T_{\text{max}}} C_{i'}(T )dT$ are reported in Table \ref{table: correlation times Reynolds dependence}, where $T_{\text{max}}$ is the duration of the largest time increment. 

\begin{table*}[!ht]
\caption{Correlation times and lengths in the streamwise direction of axial, $u'$, wall-normal, $v'$, and azimuthal, $w'$, velocity fluctuation components. The correlation times are computed as $t_{i'} = \int_0^{T_{\text{max}}} C_{i'}(T) dT$, while the correlation lengths are deduced from the zero-crossing lengths of the corresponding autocorrelation functions, $C_{i'}(X)$.}
\setlength{\tabcolsep}{1.8em}
\begin{tabular}{c c c c c c c}
\hline \\[0cm] 
$Re_\tau$ & $T_{u'}/t$ &$T_{v'}/t$ & $T_{w'}/t$ & $l_{u'}/L $ & $l_{v'}/L $ &  $l_{w'}/L $\\[0.2cm] \hline \\[0.cm]
361   & 11.092 & 17.338 & 19.318  & 0.034 & 0.188  & 0.046     \\[0.2cm]
550   & 20.379 & 22.176 & 16.622  & 0.058 & 0.033  & 0.026     \\[0.2cm]
1,000 & 18.256 & 20.099 & 20.134  & 0.052 & 0.041  & 0.077     \\[0.2cm]
2,000 & 27.731 & 33.448 & 38.431  & 0.012 & 0.122  & 0.079     \\[0.2cm]\hline
\end{tabular}
\label{table: correlation times Reynolds dependence}
\end{table*}

\newpage
\providecommand{\noopsort}[1]{}\providecommand{\singleletter}[1]{#1}%
%

\end{document}